\documentclass[conference, 10pt,letterpaper]{IEEEtran}
\usepackage[utf8]{inputenc}
\usepackage{amsmath}
\usepackage{amsfonts}
\usepackage{amssymb}
\usepackage{graphicx}
\usepackage{footnote}
\usepackage{listings}
\usepackage{subcaption}
\usepackage{cite}
\usepackage{url}

\newcommand{\rtdap}{{RT-DAP}}

\newcommand{\eat}[1]

\begin{document}

\title{\LARGE \bf {\rtdap}: A Real-Time Data Analytics Platform for Large-scale Industrial Process Monitoring and Control}

\author{
\IEEEauthorblockN{
    Song Han\IEEEauthorrefmark{1},
	Tao Gong\IEEEauthorrefmark{1},
	Mark Nixon\IEEEauthorrefmark{2},
	Eric Rotvold\IEEEauthorrefmark{2},
	Kam-yiu Lam\IEEEauthorrefmark{3},
	Krithi Ramamritham\IEEEauthorrefmark{4}\\
}
\vspace{0.08in}

\IEEEauthorblockA{
	\IEEEauthorrefmark{1} University of Connecticut, {\small \{song.han, tao.gong\}@uconn.edu}\\
	\IEEEauthorrefmark{2} Emerson Automation Solutions, {\small \{mark.nixon, eric.rotvold\}@emerson.com}\\
	\IEEEauthorrefmark{3} City University of Hong Kong, {\small cskylam@cityu.edu.hk}\\
	\IEEEauthorrefmark{4} IIT Bombay, {\small krithi@cse.iitb.ac.in}}

\eat{
\IEEEauthorblockA{
	\IEEEauthorrefmark{1} University of Connecticut, {\small \{shaobo.zheng, tao.gong, song.han\}@uconn.edu}}
\IEEEauthorblockA{
	\IEEEauthorrefmark{2} Emerson Process Management, {\small \{joshua.kidd, noel.bell, mark.nixon, eric.rotvold\}@emerson.com}}
\IEEEauthorblockA{
	\IEEEauthorrefmark{3} Microsoft, {\small \{larar, weetok\}@microsoft.com}}
\IEEEauthorblockA{
	\IEEEauthorrefmark{4} CityU of Hong Kong, {\small cskylam@cityu.edu.hk}}
\IEEEauthorblockA{
	\IEEEauthorrefmark{5} IIT Bombay, {\small krithi@cse.iitb.ac.in}}
}
}
\eat{
\author{
\alignauthor
Shaobo Zheng\titlenote{The first two authors have equal contribution to this work.}\\
       \affaddr{University of Connecticut}\\
       \email{shaobo.zheng@uconn.edu}
\alignauthor
Tao Gong\\
       \affaddr{University of Connecticut}\\
       \email{tao.gong@uconn.edu}
\alignauthor
Song Han\\
       \affaddr{University of Connecticut}\\
       \email{song.han@uconn.edu}
\and  % use '\and' if you need 'another row' of author names
\alignauthor
Mark Nixon\\
       \affaddr{Emerson}\\
       \email{mark.nixon@emerson.com}
\alignauthor
Eric Rotvold\\
       \affaddr{Emerson}\\
       \email{eric.rotvold@emerson.com}
\alignauthor
Joshua Kidd\\
       \affaddr{Emerson}\\
       \email{joshua.kidd@emerson.com}
}
\additionalauthors{
Additional authors:
Kam-yiu Lam (City University of Hong Kong, {\texttt{cskylam@cityu.edu.hk}}),
Krithi Ramamritham (IIT Bombay, {\texttt{krithi@cse.iitb.ac.in}}),
Noel Bell (Emerson Process Management, {\texttt{noel.bell@emerson.com}}),
Lara Rubbelke (Microsoft, {\texttt{larar@microsoft.com}}),
Wee Hyong Tok (Microsoft, {\texttt{weetok@microsoft.com}})}
}
\maketitle

\begin{abstract}
In most process control systems nowadays, process measurements are periodically collected and archived in historians. Analytics applications process the data, and provide results offline or in a time period that is considerably slow in comparison to the performance of the manufacturing process. Along with the proliferation of Internet-of-Things (IoT) and the introduction of ``pervasive sensors" technology in process industries, increasing number of sensors and actuators are installed in process plants for pervasive sensing and control, and the volume of produced process data is growing exponentially. To digest these data and meet the ever-growing requirements to increase production efficiency and improve product quality, there needs to be a way to both improve the performance of the analytics system and scale the system to closely monitor a much larger set of plant resources. In this paper, we present a real-time data analytics platform, called {\rtdap}, to support large-scale continuous data analytics in process industries. {\rtdap} is designed to be able to stream, store, process and visualize a large volume of real-time data flows collected from heterogeneous plant resources, and feedback to the control system and operators in a real-time manner. A prototype of the platform is implemented on Microsoft Azure. Our extensive experiments validate the design methodologies of {\rtdap} and demonstrate its efficiency in both component and system levels.
\end{abstract}

\section{Introduction} \label{sec:intro}

Key objectives in process industries include maintaining the product quality within product specifications, improving the overall efficiency of the process operations, and constantly improving health, safety and environment~\cite{luyben1989process,shinskey1990process}. To achieve these objectives practitioners have been using both first principle~\cite{blevins2010control,blevins2013advanced} and data driven methods~\cite{kadlec2009data,yin2014review}, and in some cases combinations of the two methods. In most process control systems nowadays, process measurements are periodically collected and communicated to gateways, controllers, and workstations. Feedback is provided through alarm/control messages and visualization to both the system itself and to human operators. All collected data are put into a time-series form that can be used by the modeler and archived in historians. Analytics applications read data from these historians, find the right set of features and the correct subset of data to use to capture the desired variation, and provide results off-line or in a time period that is considerably slow in comparison to the performance of the manufacturing process.

Due to the lack of an efficient and scalable real-time data analytics infrastructure specifically designed for process monitoring and control applications, only a minimal set of process conditions and plant equipment is monitored, and a limited set of control and analytics algorithms is provided. With the proliferation of Internet-of-Things (IoT)~\cite{atzori2010internet} and the introduction of ``pervasive sensors" technology in process industry, the volume of produced process measurements and the complexity of the analytics tasks are growing exponentially. To digest these data and meet the ever-growing requirements to increase production efficiency and improve product quality, there needs to be a way to provide easy access to the data, improve performance of the analytics tasks and scale the system to closely monitor a much larger set of plant resources and their operational environments.

\begin{figure}
   \centering
   \includegraphics[width=0.95\linewidth]{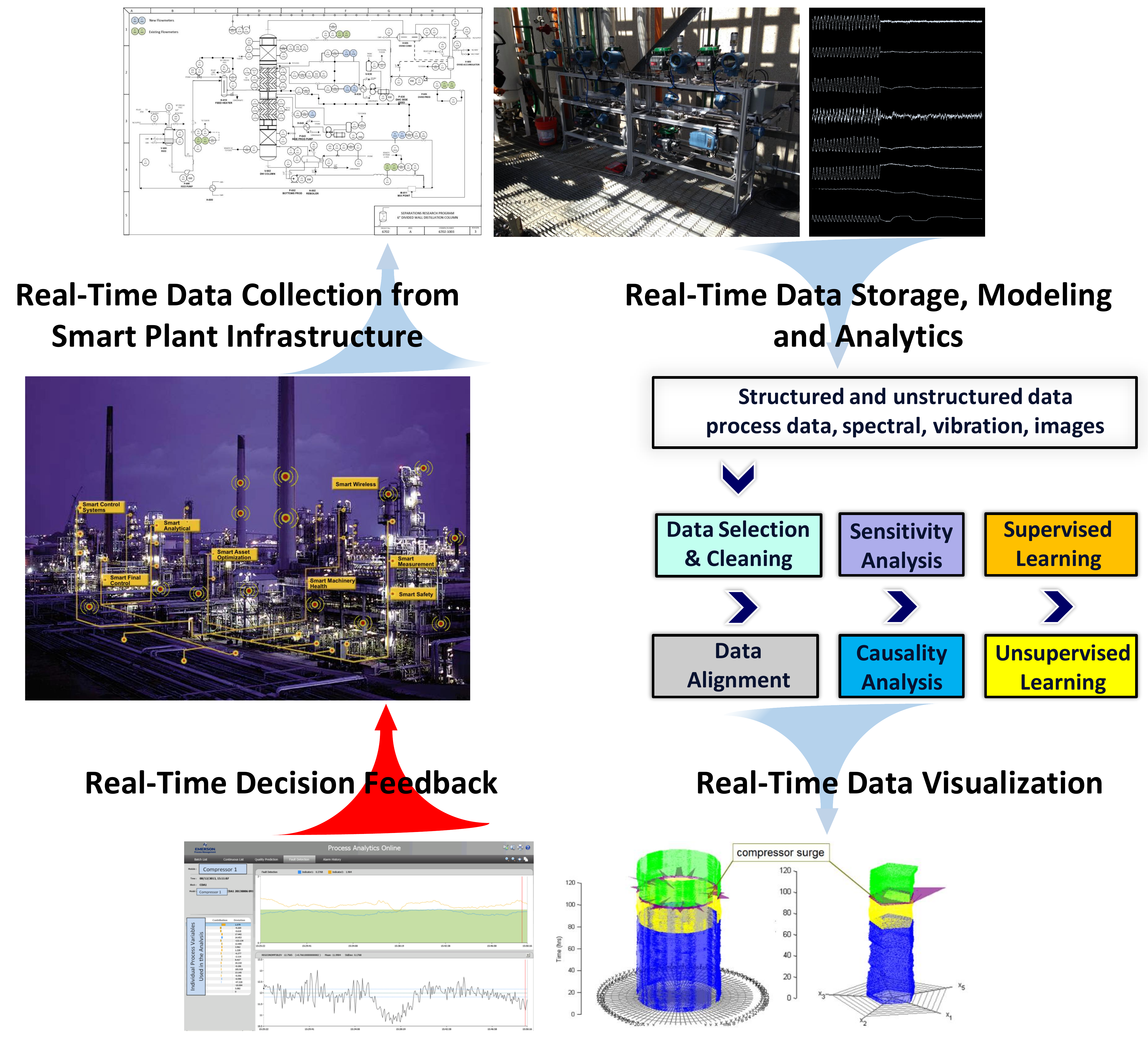}
   \caption{\small Overview of the real-time sensing, communication, analytics and control loop for next-generation process monitoring and control}
   \label{fig:overview}
\vspace{-0.2in}
\end{figure}

In this paper, we present the design and implementation of a scalable real-time data analytics platform, called {\rtdap}, to support continuous data analytics for a wide range of industrial process monitoring and control applications. As shown in Fig.~\ref{fig:overview}, {\rtdap} will serve as the core component in the real-time sensing, communication, analytics and control loop for the next-generation large-scale process monitoring and control systems. A large volume of real-time data flows will be collected from heterogenous plant resources via the smart plant communication infrastructures, typically a combination of wired network backbones and wireless edge networks. These real-time process measurements will be streamed into {\rtdap} for real-time data storage, modelling and analytics. Data visualization and control decisions will be made and fed back to the system and operators in a real-time manner for both emergency and daily process operations. \eat{The proposed platform can either be running on a private computing infrastructure or deployed on an enterprise cloud platform, such as Microsoft Azure.}

{\rtdap} differs from existing general-purpose computing platforms with the following key components: 1) a unified messaging protocol to support massive real-time process data, 2) a distributed time-series database specifically designed for storing and querying process data, and 3) a model development studio for designing data and control flows in process control related analytics tasks. The developed models and analytics modules will be deployed on an elastic real-time processing framework to distribute the computation on the parallel computing infrastructure. We further design an industrial IoT field gateway, called IIoT-FG, through which {\rtdap} can be connected to various heterogeneous plant resources for distributed data acquisition. Combined with the real-time communication infrastructure deployed in process plants nowadays, the integrated communication and computing framework will significantly improve the scalability, reliability and real-time performance of industrial process monitoring and control applications, and close the loop of process monitoring, communication, decision making, and control.

To validate the design methodologies, we implement a prototype of {\rtdap} on Microsoft Azure~\cite{azure} and a prototype of IIoT-FG on Minnowboard~\cite{minnowboard}. Our extensive experiments on both component- and system-level testing demonstrate the efficiency of the platform in providing real-time data streaming, storage, decision making and visualization for real-time analytics applications in process industries.

The remainder of the paper is organized as follows. Section~\ref{sec:related} reviews the existing real-time data analytics platforms, in particular for industrial automation applications. We present our platform design in Section~\ref{sec:design} and describe its implementation on Microsoft Azure in Section~\ref{sec:impl}. We present the performance evaluation and summarize our experimental results in Section~\ref{sec:perf}. Section~\ref{sec:concl} concludes the paper and discusses the future work.

\section{Related Works} \label{sec:related}

\begin{figure*}
\centering
\includegraphics[width=1\textwidth]{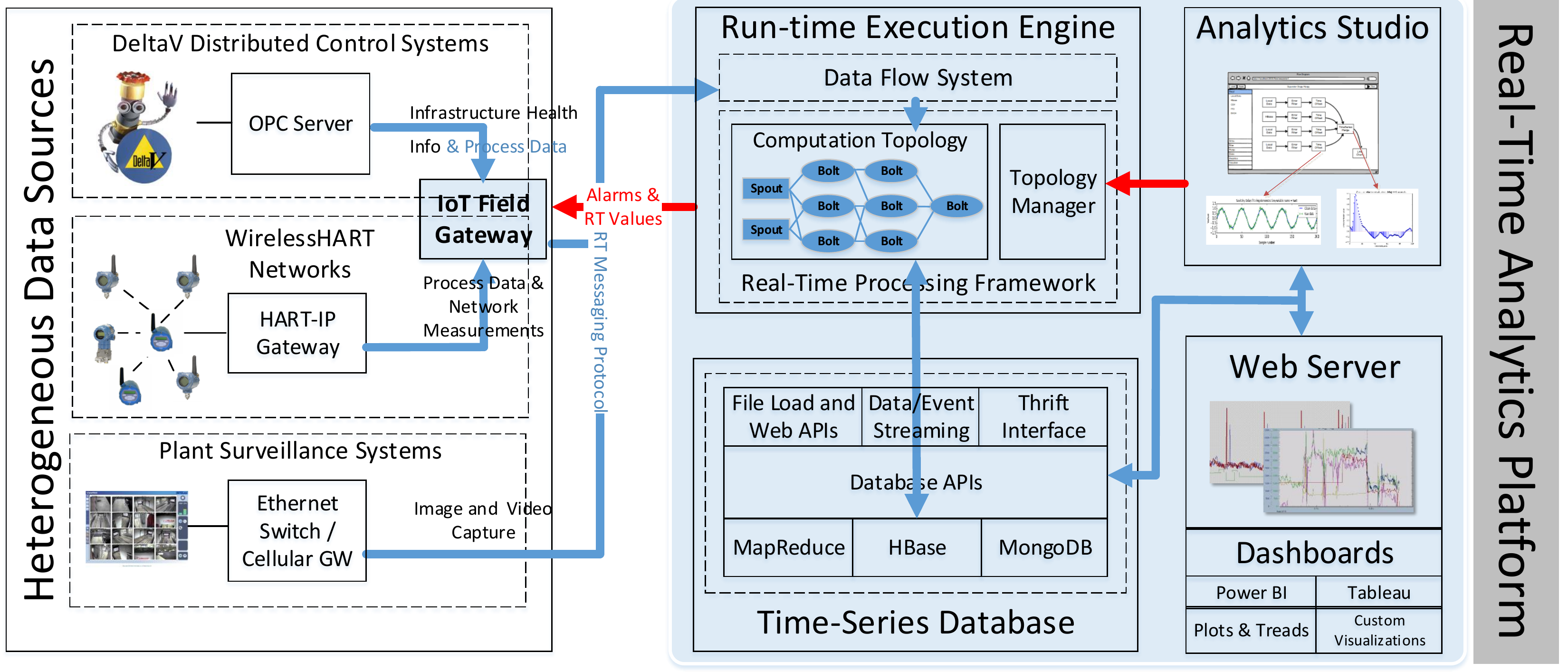}
\caption{\small An overview of the architecture of the real-time data analytics platform (RT-DAP) for large-scale process control. It has five key components: data connectors, run-time execution engine, time-series database, analytics studio and web services.}
\label{fig:arch}
\vspace{-0.1in}
\end{figure*}

Internet-of-Things (IoT)~\cite{atzori2010internet} has drawn a tremendous amount of attention in recent years. A growing number of physical objects and devices are being connected to the Internet at an unprecedented rate to collect and exchange data~\cite{al2015internet}. It is projected that 50 billion devices will be connected to the Internet by the year 2020 and the total amount of exchanged data volume will reach 35 ZB~\cite{zaslavsky2013sensing,gerhardt2012unlocking,sagiroglu2013big}. To fully utilize these data, many data acquisition and analytics systems have been developed in different application domains. In the field of industrial automation, the focus of our study in this paper, many analytics platforms have also been developed. These platforms include but are not limited to the HAVEn ~\cite{HPHAVEn} from HP, the Industrial Solutions System Consolidation Series from Intel, and the Omneo~\cite{siemens} from Siemens. Although these analytics systems are comprehensive, they are designed for general purpose industrial automation applications, and few details on their architecture design, implementation and performance evaluation are provided. In this paper, we will present the design details of {\rtdap}, which is specifically developed for industrial process monitoring and control applications. In the following, we first summarize the state of the art in the key components of the proposed platform. \eat{We then present the design details of {\rtdap} in Section~\ref{sec:design}.}

With the exponentially growing number of data sources and data volume, efficient messaging protocols and messaging systems particularly designed for IoT applications have gained their popularity in recent years. Data collected from physical objects and devices are streamed into messaging systems through messaging protocols and eventually consumed by IoT applications. Among many existing messaging protocols, MQTT~\cite{mqtt}, AMQP~\cite{vinoski2006advanced}, CoAP~\cite{mqttcoapIoT} and STOMP~\cite{stomp} are the four most popular ones based on TCP/IP. All these protocols are designed for resource constrained devices and networks, thus can be efficiently implemented on a variety of embedded systems. A comparison among these popular IoT messaging protocols are provided in~\cite{msgprotocolcomparison}.

\eat{
MQ telemetry transport (MQTT) is a machine-to-machine (M2M) IoT connectivity protocol~\cite{mqtt}. MQTT enables the publish-subscribe mode and provides many-to-many broadcast capabilities. Although light weight is a common feature in all IoT messaging protocols, MQTT is extremely light weight that it has only five methods. This design works perfectly on small, low power devices. On the other hand, the lack of functionality in MQTT may not satisfy larger systems' requirements.

Another widely used messaging protocol is Advanced Message Queuing Protocol (AMQP). AMQP is an open Internet protocol for reliably sending and receiving messages~\cite{vinoski2006advanced}. AMQP makes it possible for everyone to build a diverse, coherent messaging ecosystem capable for a wide range of features related to messaging, including reliable queuing, topic-based publish/subscribe messagging, flexible routing, transactions and security~\cite{qpid2013open,amqpvmware}.

The Constrained Application Protocol (CoAP) is a specialized web transfer protocol intended for use with resource-constrained nodes and networks that need to be controlled or supervised remotely. CoAP packets are much smaller than HTTP TCP flows for saving space. Unlike MQTT, CoAP is primarily an one-to-one protocol for transferring state information between client and server~\cite{mqttcoapIoT}.

STOMP (Simple/Streaming Text Oriented Messaging Protocol) is a text-based protocol, analogous to HTTP. Similar to AMQP, STOMP provides a message (or frame) header with properties, and a frame body. The design principles are to create something simple, and widely-interoperable~\cite{amqpvmware}.}

In order to partition and separate the data streams, and process messages asynchronously, most messaging systems are carefully designed with features like scalability, reliability, clustering, multi-protocol and multi-language support. Messaging systems usually comprise several servers, known as messaging brokers, working in the middle of the data sources and data analytics systems. These systems such as ActiveMQ~\cite{ActiveMQ}, RabbitMQ~\cite{rabbitmq}, ZeroMQ~\cite{zeromq}, and Kafka~\cite{kreps2011kafka} share many similarities. Some throughput and latency benchmarks are provided in~\cite{braveNewGeek} for their performance comparison.

\eat{
ActiveMQ implements the Java Message Service (JMS) specification, and is an open source messaging and Integration Patterns server. It provides full support for the Enterprise Integration Patterns both in the JMS client and the Message Broker~\cite{ActiveMQ}. RabbitMQ is one of the open source softwares that support several messaging protocols including AMQP, STOMP, MQTT and HTTP through the use of plugins. RabbitMQ supports both persistent and non-persistent delivery. To guarantee delivery, the brokers use message acknowledgements which incur a massive latency penalty~\cite{braveNewGeek}.

Unlike other messaging systems, no message brokers are needed in the implementation of ZeroMQ and this leads to a higher throughput and lower latency. The drawbacks of such design is also explicit: though ZeroMQ guarantees that messages will be delivered atomically intact and ordered but it does not guarantee the delivery of them.

Kafka is a distributed messaging system for collecting and delivering high volumes of log data with low latency~\cite{kreps2011kafka}. Also, it is suitable for both offline and online message consumption. High availability and horizontally scaling are also other key features of Kafka through the use of ZooKeeper.}

Data store is another critical component of a data analytics system. Relational databases have been widely used in past decades, but it is not suitable for storing a large amount of time series data without explicit structures and relations~\cite{madden2012databases}. A variety of NoSQL databases thus have been designed to address this problem, including key-value stores and document stores. Key-value stores -- including Berkeley DB (BDB)~\cite{olson1999berkeley}, Oracle NoSQL Database~\cite{joshi2012oracle}, HBase~\cite{hbase}, and Cassandra~\cite{cassandra} -- use a map or dictionary as a collection of key-value pairs as the fundamental data model. On the other hand, Document stores differ on the definition of ``document", but generally all assume that documents encapsulate and encode data in some standard formats or encodings including XML, YAML~\cite{yaml}, and JSON. CouchDB~\cite{couchDB} and MongoDB~\cite{mongoDB} are among the most popular Document stores. \eat{In our proposed data analytics platform, since the majority of the data are time series process measurements, we chose HBase for storing the raw data given the need of key-based data access when storing and retrieving the sensor measurements.}

Regarding the parallel computing framework, MapReduce~\cite{mapreduce}, Storm~\cite{storm} and Spark~\cite{engle2012shark} are the three most widely employed computation models nowadays for big data processing, while they have different strengths and use cases. MapReduce is a software framework capable for processing vast amounts of data (multi-terabyte datasets) in parallel on large clusters of commodity hardware in a reliable manner with high fault-tolerance~\cite{HadoopMapReduce,dean2008mapreduce}. Apache Storm is a distributed computation framework which is suitable for reliably processing unbounded data streams. It uses custom created ``spouts" and ``bolts" to define information sources and manipulations to allow batch, distributed processing of streaming data. In Storm, all data are stored in memory if the message exchanging happens in the same machine. This design makes Storm fast enough to process huge amount of data in real-time. Apache Spark is an open source cluster computing framework. In contrast to Hadoop's two-stage disk-based MapReduce paradigm, Spark's multi-stage in-memory primitives provide performance up to 100 times faster for certain applications. By allowing user programs to load data into a cluster's memory and query it repeatedly, Spark is well-suited to machine learning algorithms~\cite{engle2012shark}. A detailed design overview and performance comparison between the Storm and Spark streaming platforms is provided in~\cite{StormSparkCompare,StormSpark}. It shows that Storm is more suitable for stateless stream processing while the Spark is more useful in complex event processing.

\begin{figure}
\includegraphics[width=\columnwidth]{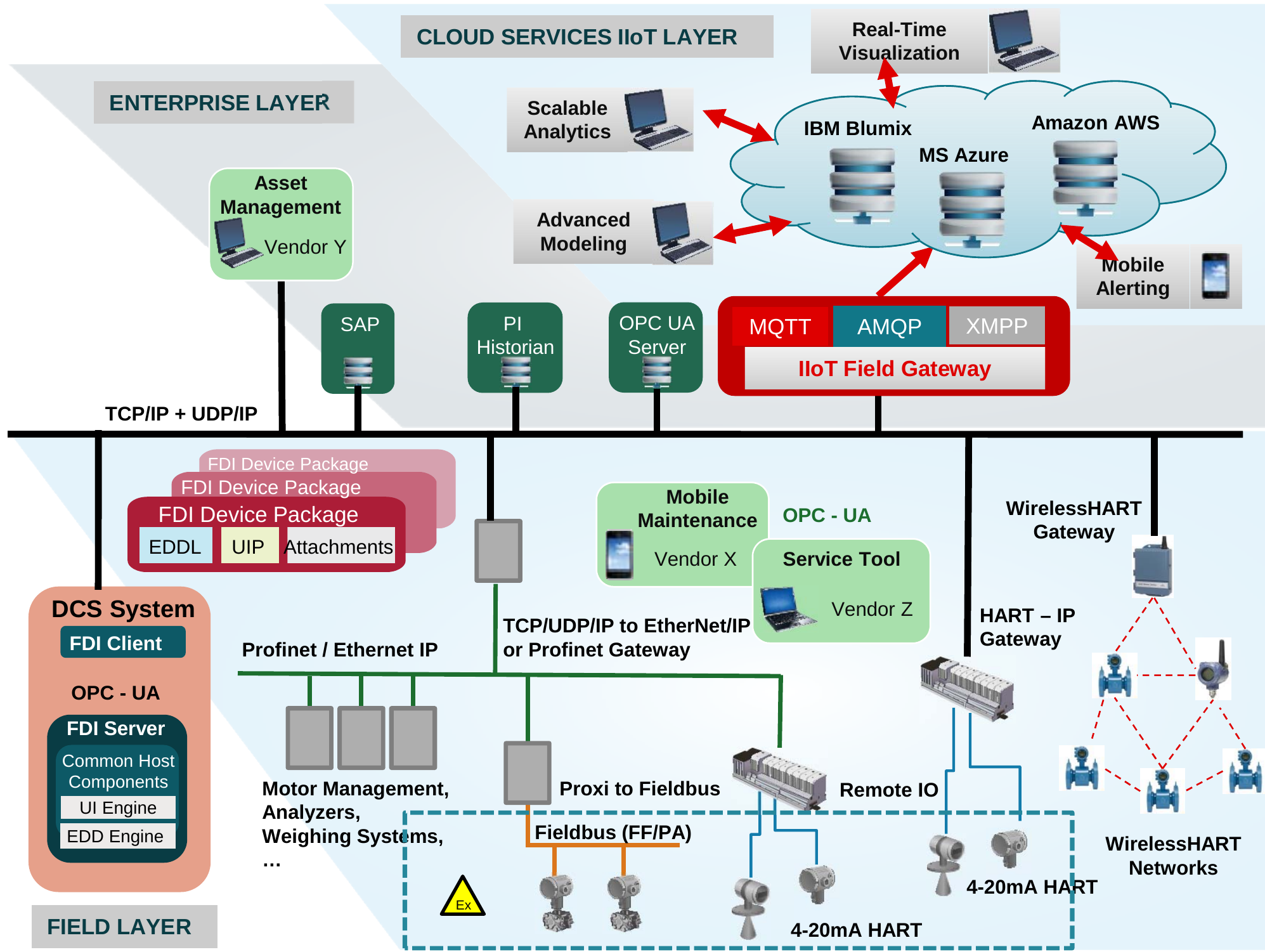}
\caption{\small Vision of the Industrial IoT in process industries}
\label{fig:vision-iiot}
\vspace{-0.2in}
\end{figure}

\section{{\rtdap} Design Details} \label{sec:design}

In this section, we present the design details of {\rtdap}. {\rtdap} is a synergy of multiple advanced communication and computing technologies and is expected to provide a scalable solution for large-scale real-time industrial process monitoring, control and analytics. Fig.~\ref{fig:arch} presents the overall architecture of {\rtdap} which consists of five key components: one or multiple industrial IoT field gateways for connecting to heterogenous plant resources, a run-time execution engine for real-time data processing, a distributed time-series database for fast data loading and queries, an analytics studio to define analytics models, and a rich set of web services for real-time data visualization and interactions. The industrial IoT field gateway, called IIoT-FG, provides hardware interfaces and protocol adaptation between RT-DAP and various heterogeneous plant resources running different communication protocols (OPC-UA~\cite{opcua}, HART-IP~\cite{hart-ip}, etc.) for distributed data acquisition. The collected process measurements from the field are in essence time-series data, and will be streamed into {\rtdap} through a unified messaging protocol. These real-time data will be digested in the run-time execution engine which is a combination of data flow system (like Kafka~\cite{kafka}) and parallel real-time computing framework (like Storm~\cite{storm}) where data analytics tasks will be performed in a parallel and resource-aware manner. The computed results will either be fed back to the physical systems directly in the forms of notification and alarm messages, or loaded into the time-series database system for further queries and processing. Another important component, the analytics studio is designed to develop various anayltics models using first principle and/or data-driven methods. These developed models will be deployed in the run-time execution engine for performing online continuous data analytics. Lastly, a web server is developed to interact with the time-series database to provide real-time visualization to end users through a variety of dashboards. In the following, we will elaborate the design of each key components.

\subsection{Naming Conventions} \label{sec:design:preliminaries}

We first describe our naming convention to distinguish real-time data points from different plant resources. The format is consistent with the one used in DeltaV~\cite{deltav} system, and other major Distributed Control System (DCS) vendors apply a similar approach. Each data point is assigned a unique tag name. Within one zone, the top name of the tag can be one of three types: module, workstation/controller, and device. The data points are all defined as paths from the top name. In a plant of multiple zones, zone name will be prefixed to the top name. To further distinguish different plants, domain name will be prefixed to the zone name. For ease of presentation, in this paper we only consider the data points from a single plant and thus domain name is not included. Considering an example where we have a zone named ``UCONN-ITEB-311". Inside this zone, there is one wireless Gateway with a 5-byte UniqueID of ``A5EF69D256", which has one sensor device connected with a 5-byte UniqueID of ``A286BD21FA". In the following, Tag-1 represents the health status of the Gateway, and tag-2 and tag-3 represent the health status and primary variable (PV) output value of the sensor device respectively. \\

\vspace{-0.05in}
\noindent {\small Tag-1: UCONN-ITEB-311::A5EF69D256/Health} \\
\noindent {\small Tag-2: UCONN-ITEB-311::A5EF69D256/A286BD21FA/Health} \\
\noindent {\small Tag-3: UCONN-ITEB-311::A5EF69D256/A286BD21FA/OUT.PV}

\subsection{Industrial IoT Field Gateway} \label{sec:design:dau}

Fig.~\ref{fig:vision-iiot} presents our vision of the industrial IoT paradigm in process industry. A large number of real-time data points from heterogeneous plant resources will be collected from a variety of data connectors (both hardware devices and software interfaces) which are geographically distributed in the field. For example, OPC UA servers are used to connect to DeltaV systems to retrieve periodic measurements from installed modules, controllers and hardware devices. HART and WirelessHART gateways are used to collect sensor and actuator measurements as well as network health information in a real-time and continuous manner. Wireless packet sniffers, spectrum analyzers and surveillance cameras (not shown in Fig.~\ref{fig:vision-iiot}) are installed in the plant to monitor its operation and RF spectrum environments. All these real-time data will be streamed into cloud-based data analytics platform(s) (such as {\rtdap}) for advanced modeling, scalable analytics, real-time visualization and mobile alerting.

\begin{figure}
\includegraphics[width=\columnwidth]{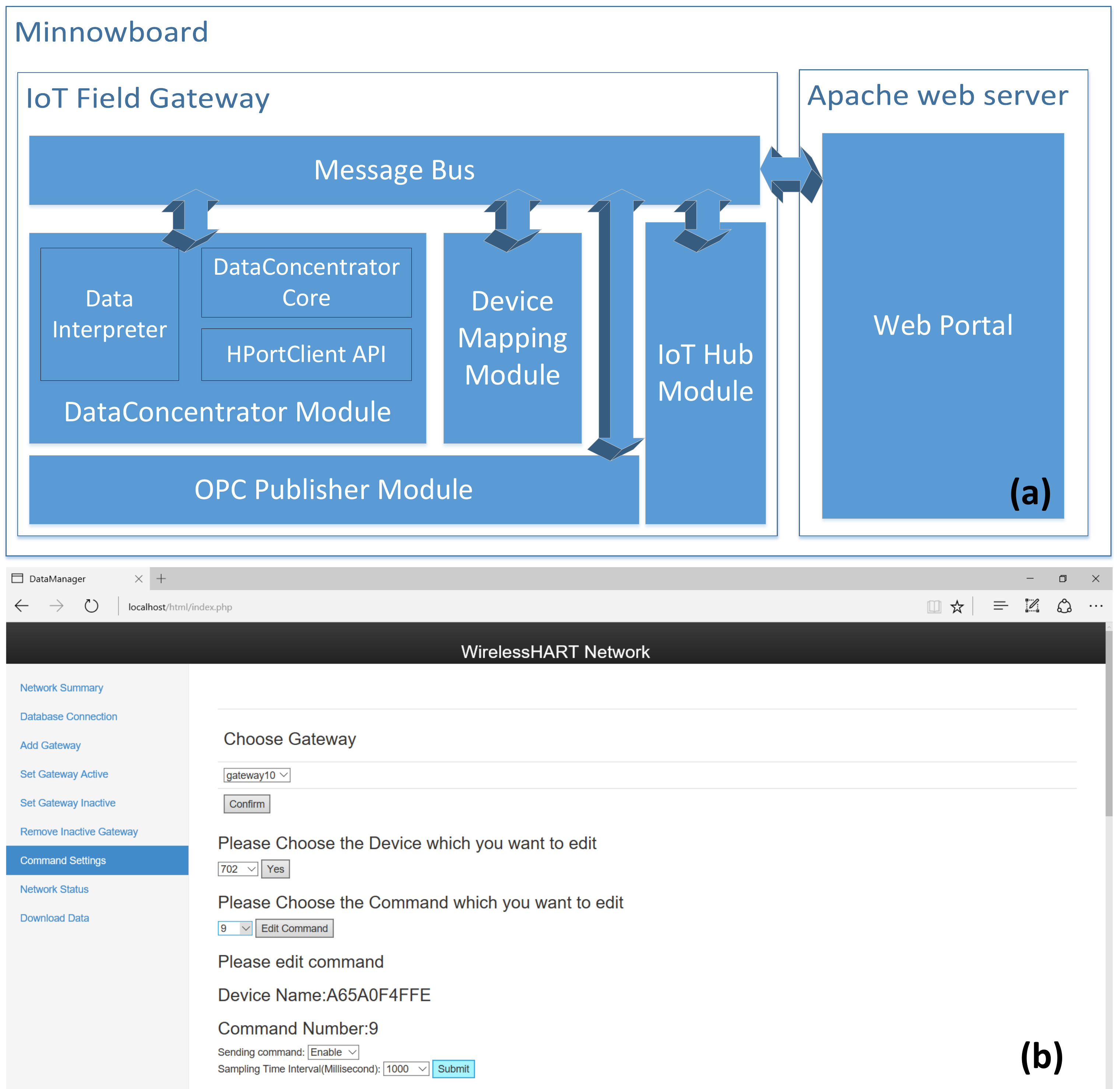}
\caption{\small (a) Software architecture of the IIoT field gateway (IIoT-FG); (b) web portal for remote access and network/device configuration.}
\label{fig:DataConcentratorOverview}
\vspace{-0.2in}
\end{figure}

Given the connectors are usually running different communication protocols ({\em e.g.}, OPC UA and HART-IP), instead of implementing protocol adapters on each of the connectors to stream the data to the cloud, we design an industrial IoT field gateway, referred to as IIoT-FG, to connect to multiple data connectors to provide protocol adaptation and remote configuration. IIoT-FG is designed to be of small form factor, cheap, and thus can support massive field deployment.

\eat{
\begin{itemize}
\item Remote configuration and data streaming: The data concentrator can be easily connected to WirelessHART Gateways and automatically detect all operational devices through HART-IP~\cite{hart-ip} protocol. By specifying the data points and associated parameters through a web portal, users can stream the measurements into the analytics platform in a real-time manner.
\item Dynamic network statistics and visualization: The data concentrator can display the dynamic network topologies and summarize various network/device statistics.
\item Data caching: The data concentrator can cache the network health data and sensor measurements for a specified time period. This is to mitigate packet loss in the presence of network failure and power shutdown.
\end{itemize}}

Fig.~\ref{fig:DataConcentratorOverview}(a) presents the software architecture of IIoT-FG. The current prototype is implemented on Minnowboard and has the following major software modules: (1) a web portal running on Apache to enable remote configuration, (2) a dataconcentrator module and an OPC publisher module to interpret HART-IP and OPC UA messages respectively, (3) a device mapping module to map device UID to the device key used on the analytics platform, and (4) an IoT hub module to stream the data to the data analytics platform by supporting different messaging protocols, such as HTTP, MQTT and AMQP. All these modules communicate with each other through a message bus. As shown in Fig.~\ref{fig:DataConcentratorOverview}(b), operators can remotely access to the IIoT-FG via the web portal and configure the target data points and the associated streaming parameters. Process measurements will be streamed into IIoT-FG for protocol adaptation, and then further forwarded to the data analytics platform according to the messaging protocol to be described in Section~\ref{sec:design:streaming}. In the data analytics platform, these data points will be further stored, fused, analyzed and visualized to represent the current status of plant operations.

\subsection{Messaging Protocol for Data Collection} \label{sec:design:streaming}

Along with the ever-growing number of sensors and actuators being deployed in field, a large amount of real-time measurements are being collected from heterogeneous plant resources for various monitoring and control applications. A simple and unified streaming protocol is thus needed to define these data streams for cross-platform data emitting and retrieving. Given its capability to represent rich data structures in an extendable way, we use JSON objects to define these data streams. In {\rtdap}, we set up a TCP portal server based on Netty and design a streaming API for clients to interact with the TCP server. This streaming API is designed with two fields at the top level: the request \emph{type}, and its associated \emph{parameter}, which is another JSON object with multiple fields. The followings are two major request types:

\eat{
\vspace{0.02in}
\noindent {\em Stream Definition:} is created by the client to define the data stream before sending any data records to the server. This request has a type of ``D" and an associated {\em parameter} with \emph{id}, \emph{tag}, and \emph{type} fields. This is to indicate that the stream has a unique \emph{id} and is mapped to the given \emph{tag} name. Upon receiving such request, the server will create the mapping.

\vspace{0.02in}
\noindent {\em Data Record:} After a stream is defined, the client can emit data records through the Data Record request. This request has a type of ``d" and an associated parameter with fields of \emph{id}, \emph{time}, \emph{value}, and \emph{status}. This is to represent a data record from stream \emph{id}, with its timestamp, data value and status.
\vspace{0.02in}}

\begin{itemize}
\item {\em Stream Definition:} This request type is created by the client to define the data stream before sending any data records to the server. It has a type of ``D" and a {\em parameter} with \emph{id}, \emph{tag}, \emph{type} and optional fields. This indicates that the stream has a unique \emph{id} and is mapped to the \emph{tag} name. Upon receiving such a request, the server will create the mapping.

\item {\em Data Record:} After a stream is defined, the client can emit data records through the Data Record request. This request has a type of ``d" and an associated parameter with fields of \emph{id}, \emph{time}, \emph{value}, and \emph{status}. This is to represent a data record from stream \emph{id}, with its timestamp (in UTC format), data value and status.
\end{itemize}

\eat{Fig.~\ref{fig:streaming-format} gives two example JSON objects for the streaming API. The Streaming Definition request in Fig.~\ref{fig:streaming-format}(a) creates a stream with an {\em id} of 23 and maps it to a {\em tag} with name of ``zone::gateway/device/OUT.CV". Its data type is float (F). The Data Record request in Fig.~\ref{fig:streaming-format}(b) submits a data record sampled at timestamp 1380028338000 (Sep.24 2013 13:12:18.000), with a value of 24.75 and status of 128.}

Given the fact that JSON essentially sends its schema along with every message, it requires relatively large bandwidth. Since many compatible compression techniques have been reported to achieve good JSON format compression rates, they can be performed on Data Record requests to make the streaming protocol more bandwidth-efficient. Our performance evaluation in Section~\ref{sec:perf:tcp} summarizes our experimental results on how the compression techniques will affect the throughput of the TCP portal server. All data records collected through the streaming APIs will be streamed into {\rtdap}.

\subsection{Scalable Time Series Database Design} \label{sec:design:database}

\begin{figure*}
\includegraphics[width=1.0\textwidth]{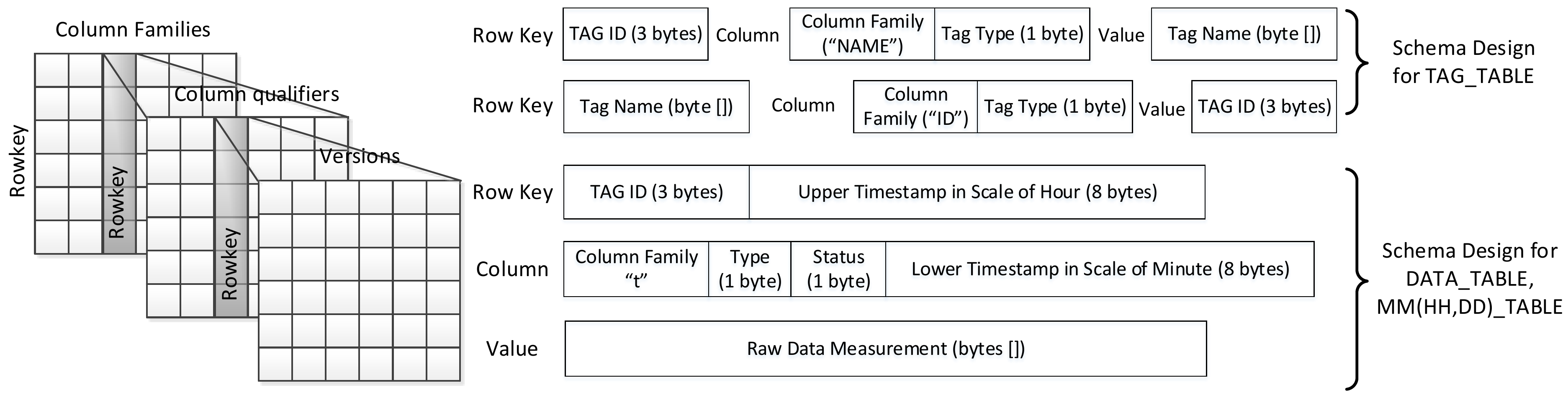}
\caption{\small The database schema design for tag, raw data and aggregation data tables in the time series database.}
\label{fig:HBase-4D}
\vspace{-0.1in}
\end{figure*}

Data records collected from plant resources include continuous, batch, event, and other data sources such as lab systems and material handling systems. These data records are in essence time series data and need to be periodically transferred to the system's Real-Time Database (RTDB) to provide a complete picture of the plant operation and support operator trends, process analysis, model building and data mining activities. To serve these purposes, we design a distributed and scalable time series database on top of HBase~\cite{hbase}. The time series database addresses a common need: store, index and serve process and related data collected from the distributed control systems (control strategies, DCS equipment, devices, lab systems, applications, etc.) at a large scale, and make this data easily accessible. We design the time series database as a general-purpose data store with a flexible data model. This allows it to craft an efficient and relatively customized schema for storing its data.

The non-relational database mechanisms in HBase enable design simplicity, horizontal scaling, and finer control over data availability. In the logical data model of HBase as shown in Fig.~\ref{fig:HBase-4D}, it stores a piece of data within a table based on a 4D coordinate system: rowkey, column family, column qualifier, and version. In our design, the rowkey of the raw data table (DATA\_TABLE) consists of the tagID of the data stream and higher-order of the timestamp when a data record is received; the column qualifier contains the data type, status and the lower-order of the timestamp; the column family is reserved for future use, and the size and contents of the value field depends on specific data streams. We have this design to sort the time series according to their unique names and time resolutions, so that the set of values for a single TagID is stored as a contiguous row. Within the run of rows for a TagID, stored values are ordered by timestamp. The timestamp in the rowkey is rounded down to the nearest 60 minutes so a single row stores a `bucket' of measurements for the hour. Dividing the rowkey this way allows users to shrink their scan range by specifying the names of the target data streams and the required time intervals.

\eat{
\begin{figure}
\centering
\includegraphics[width=0.4\textwidth]{figs/streaming-protocol.pdf}
\caption{\small Two example JSON objects for defining (a) Streaming Definition request and (b) Data Record request.}
\label{fig:streaming-format}
\vspace{-0.2in}
\end{figure}}

Following the design principles of HBase, our database design installs all raw data records into one big table (DATA\_TABLE). The table can be automatically split and distributed to multiple region servers in HBase for fault tolerance and scalability. In addition to the raw data table, as shown in Fig.~\ref{fig:HBase-4D}, we create an index table (TAG\_TABLE) which contains two column families: `tag name' and `id'. This table provides a 2-way dictionary of tag name to TagID. Based on the requirements of typical process control applications, we further create a set of aggregation tables (MM\_AGG\_TABLE, HH\_AGG\_TABLE, DD\_AGG\_TABLE) to store aggregated data  about the rows according to different time resolutions, for example minimum, maximum and close values for a given tag in each hour. This allows discarding time ranges that are known not to include data in the search range without scanning each sample. The creation and update of the aggregation tables can either be done in the runtime or through executing offline MapReduce jobs.

\eat{
\begin{figure}
\includegraphics[width=0.48\textwidth]{figs/TAG_TABLE.png}
\caption{\footnotesize Schema Design for Table TAG\_TABLE}
\label{fig:TAG_TABLE}
\end{figure}

\begin{figure}
\includegraphics[width=0.48\textwidth]{figs/DATA_TABLE.png}
\caption{\footnotesize Schema Design for Table DATA\_TABLE}
\label{fig:DATA_TABLE}
\end{figure}

\begin{figure}
\includegraphics[width=0.48\textwidth]{figs/HealthDataTable.png}
\caption{\footnotesize Schema Design for Inserting Network Health Data to Table DATA\_TABLE}
\label{fig:NetworkHealthData}
\end{figure}}

\eat{\vspace{0.1in}
\noindent {\bf TODOs: summarize the MongoDB design and its integration with HBase. Discuss what raw data and meta data we will store and is our current approach and database schema sufficient?}}

\subsection{Analytics Model Development Studio} \label{sec:design:model}

\eat{
\begin{figure*}[t!]
    \begin{subfigure}[t!]{0.32\textwidth}
        \centering
        \includegraphics[width=1\textwidth]{figs/MDS1.png}
		\caption{Environment}
		\label{fig:MDS1}
    \end{subfigure}
    ~
    \begin{subfigure}[t!]{0.32\textwidth}
        \centering
		\includegraphics[width=1\textwidth]{figs/MDS2.png}
		\caption{Model}
		\label{fig:MDS2}
    \end{subfigure}
    ~
    \begin{subfigure}[t!]{0.32\textwidth}
        \centering
		\includegraphics[width=1\textwidth]{figs/MDS3.png}
		\caption{Online view}
		\label{fig:MDS3}
    \end{subfigure}
\caption{Model Development Studio}
\end{figure*}}

Developing an analytics-based solution not only requires access to data, but also an environment to develop models using that data, an easy method to deploy models, and a way to monitor the operation of the deployed models. For example in process industry, the overall scope of a project may be to ensure that the separations coming out of a column are maintained across a wide range of material compositions. The way to do this is to provide continuous and timely feedback on how the separations column is performing. In many cases it may be possible to install and utilize one or more on-line analyzers. However, in some cases it may be too expensive or not possible to install an analyzer in the process. In other cases, there may not be an analyzer available to measure the properties that are needed for the control strategy. As an alternative inferred measurement may be used. Each inferred measurement utilizes multiple parameters. Each inferred measurement may be periodically validated and the models recalibrated using lab data. Each of these inferred measurements is developed using our self-developed tool called Model Development Studio (MDS).

MDS provides a visual workspace to build, test, deploy, and monitor analytics strategies. The studio environment makes it easy for the model developer to develop and test using different datasets, different data cleaning techniques, and different algorithms. The environment is both interactive and visual. The user creates an analytics module and work inside it. The analytics strategy inside an analytics module is constructed from a pallet of analytics blocks. Blocks are arranged into categories. Each category is used to hold blocks for accessing data, cleaning data, manipulating data, modeling, and testing.

\begin{figure}
    \vspace{0.1in}
    \centering
    \includegraphics[width=0.48\textwidth]{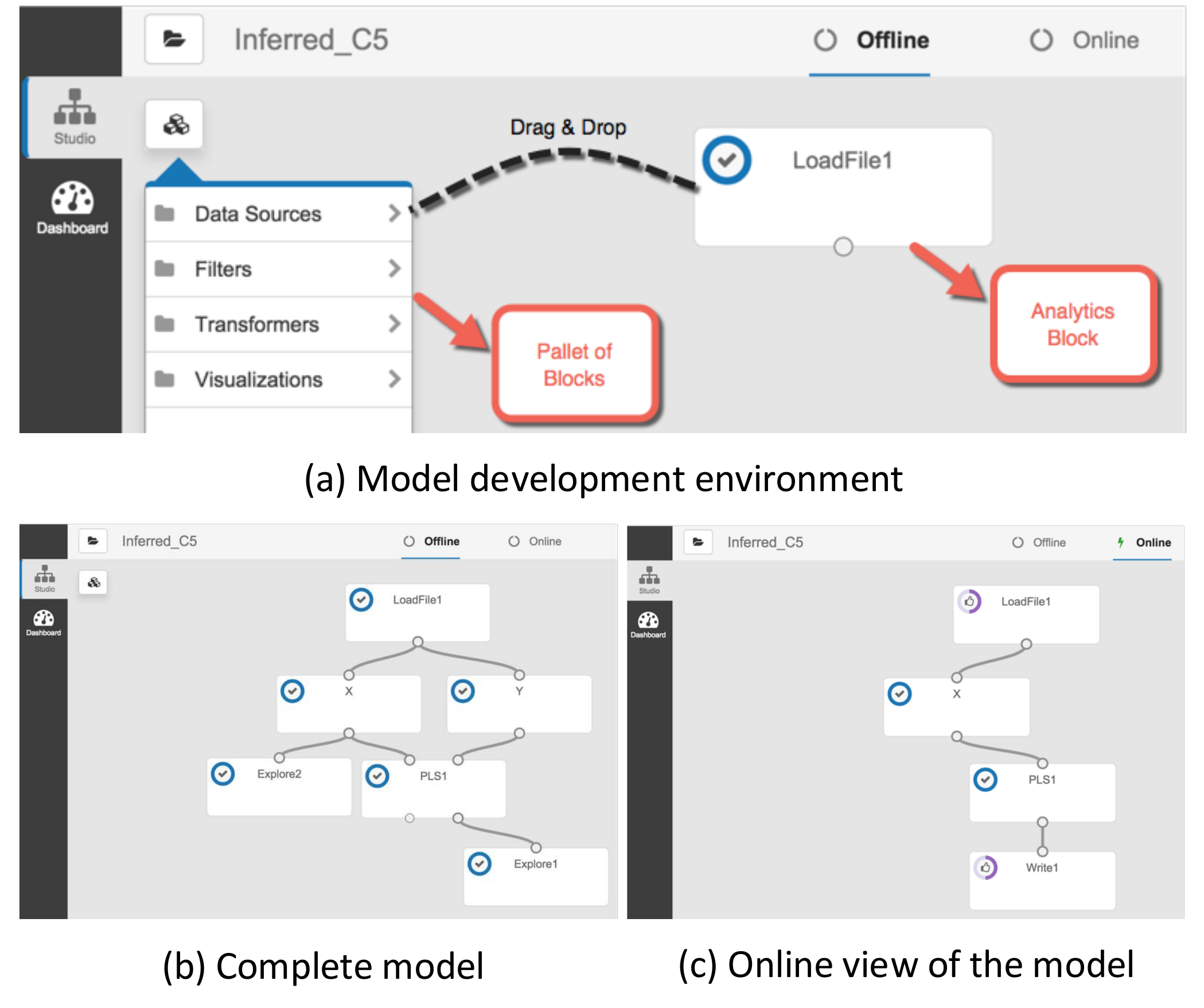}
	\caption{\small An example of model development studio (MDS)}
    \label{fig:MDS}
    \vspace{-0.2in}
\end{figure}

An MDS example is shown in Fig.~\ref{fig:MDS}(a). In this example the analytics module has one instantiated analytics block called \emph{LoadFile1}. As a first step the data could be loaded into MDS and visualized. This first step is often used to get a quick feel for shape of the data and to summarize missing data, outliers, and bad data. To complete the inferred measurement described above the data must be separated into features used to predict the measurement we are after. The completed model is shown in Fig.~\ref{fig:MDS}(b). Once the analytics strategy is ready it may be deployed for online operation using the Online option. The online operation automatically strips off blocks that are needed for runtime execution of the model. The online view also provides options for the user to control execution of the model and direct the output of the block, for example the inferred measurement may be written back to the control system using OPC or other interfaces. The online view is shown in Fig.~\ref{fig:MDS}(c).

\begin{figure}
\centering
\includegraphics[width=0.43\textwidth]{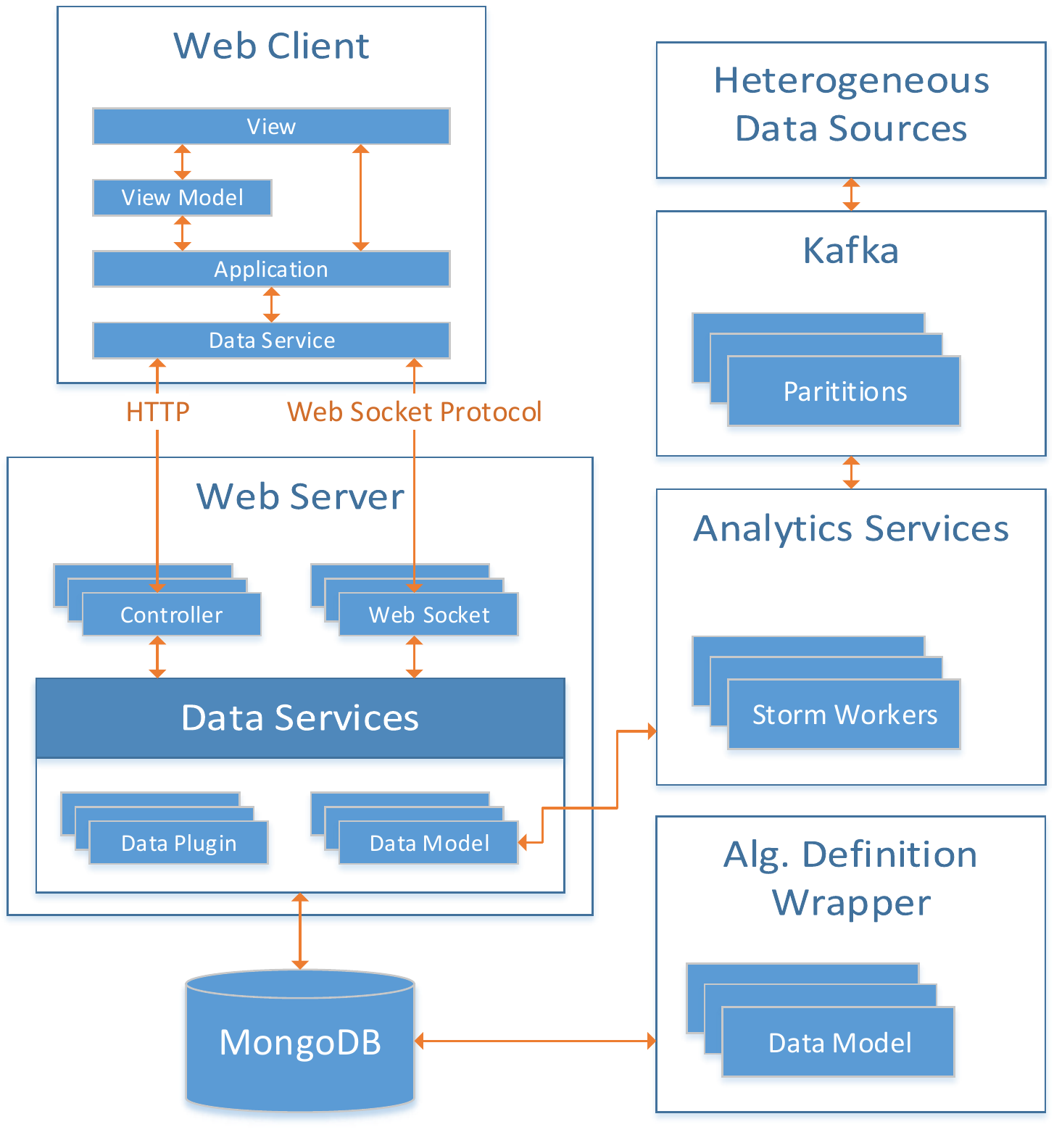}
\caption{\small Overall architecture of the model development studio}
\label{fig:MDS-overall}
\vspace{-0.2in}
\end{figure}

The overall architecture of MDS is shown in Fig.~\ref{fig:MDS-overall}. Its front-end is a web-based application. The studio environment itself contains a menu structure for creating and deploying modules, a side-bar for switching between studio and a run-time dashboard, a pallet for organizing analytics blocks, and a work surface for organizing the data flow through blocks. The model editor also supports dragging/dropping blocks from the pallet of blocks on to the work surface and connecting blocks on the work surface.

The Web Server contains a controller for interacting with the client and web socket interface for serving up data. The actual analytics blocks themselves are designed separately and loaded into the analytics environment. Each block contains both an algorithm and a set of meta data describing the interfaces for the block. In this way the model clients don't need to know anything about the internals of blocks.

Analytics blocks are defined externally using a wrapper, which describes both the offline aspects for off-line editing and on-line operation. Block definitions are defined in the wrapper and then loaded into a MongoDB. Once in the database the blocks are immediately available to the editor. When users drag blocks on to the work surface they are instantiating an analytics block. The structure and configuration of each analytics model is stored in the MongoDB as a separate entity.

\subsection{Real-Time Runtime Execution Engine} \label{sec:design:processing}

The developed models in the MDS will be deployed on the real-time runtime execution engine to perform designated analytics tasks. Among many existing computing frameworks, the MapReduce provides good performance in processing large datasets with a parallel, distributed algorithm on a cluster, and brings in scalability and fault-tolerance by optimizing the execution engine once. Our proposed {\rtdap} fully supports running complex MapReduce jobs on HBase datasets to perform computation intensive analytics tasks for process monitoring and control. MapReduce however is not a good choice for processing unbounded real-time data streams. It is hard to achieve processing rates with short latencies, which is critical for real-time continuous analytics. Although we can run the MapReduce jobs periodically, the startup and shutdown cost on a MapReduce job is proved to be heavy.

In {\rtdap}, we use a combination of Apache Kafka~\cite{kafka} and Storm~\cite{storm} frameworks to serve as the runtime execution engine. The real-time data measurements streamed to the portal server will be pushed to the Kafka framework for queuing and to achieve ``at least once" delivery guarantee. These measurements will then be pulled by the Storm framework for real-time and parallel processing. Storm provides an enhanced computation model by extending MapReduce jobs to a computation topology. Incoming data streams are split among a number of processing pipelines. Each node on the computation topology run a specific job continuously on the unbounded data streams flowing through it by using long-lived processes, and thus amortize the startup costs to zero and significantly improve the latency.

\begin{figure}
\centering
\includegraphics[width=0.38\textwidth]{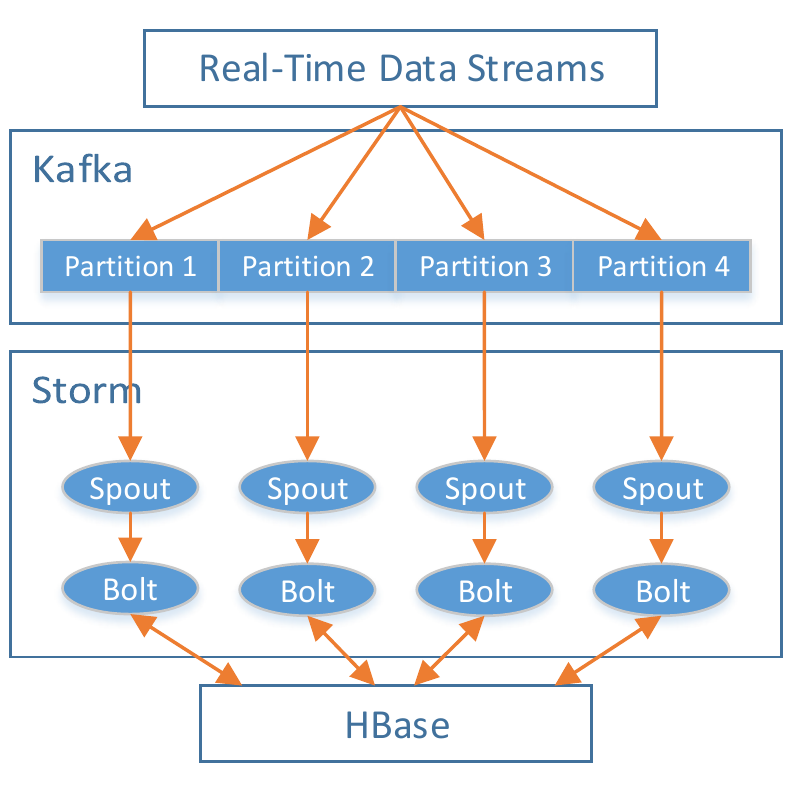}
\vspace{0.05in}
\caption{\small The Kafka partition and Storm topology for the parallel data aggregation task.}
\label{fig:storm-example}
\vspace{-0.2in}
\end{figure}

Fig.~\ref{fig:storm-example} gives an example of Kafka partition and Storm topology to support parallel data aggregation tasks. Real-time data records are streamed into different Kafka partitions according to their tag IDs. The data records in each Kafka partition are handled by a separate Storm topology which contains one Spout and one Bolt. The Spout subscribes to the corresponding Kafka partition and keeps pulling in the data records and stores them in a local buffer. The Bolt receives the buffered data records from the Spout, accesses the HBase to retrieve historical data, calculate the min, max and close values according to different time resolutions (minute, hour and day), and update the new aggregation results back to HBase.

Kafka and Storm have been shown in our experiments to be very effective in queuing and processing unbounded real-time streams. It is however difficult for the system designer to decide how to create and optimize the computation topology so that the timing constraints on the analytics jobs can be met. To overcome these deficits, we are working on an enhanced computation model for the existing real-time processing frameworks by taking timing requirements of the analytics jobs into consideration. As the ongoing work, we are exploring how to automatize the parallelizing process on the existing analytics model into a computation topology. Another challenge is how to dynamically allocate physical computing resources to each computation unit in the computation topology, so that the overall required computing resources will be minimized while the timing constraints on the analytics tasks can still be maintained.

\eat{\vspace{0.1in}
\noindent {\bf TODO: Discuss the elastic real-time resource allocation with hard or soft real-time constraints on the analytics tasks. Give a concrete example (data aggregation) to show how Kafka and Storm are integrated together.}}

\section{Implementation on Azure}
\label{sec:impl}

{\rtdap} can either run on a private computing infrastructure or be deployed on an enterprise cloud platform, such as Microsoft Azure. In this section we describe the  implementation details of our prototype development on Microsoft Azure.

\begin{figure}
   \centering
   \includegraphics[width=1.0\linewidth]{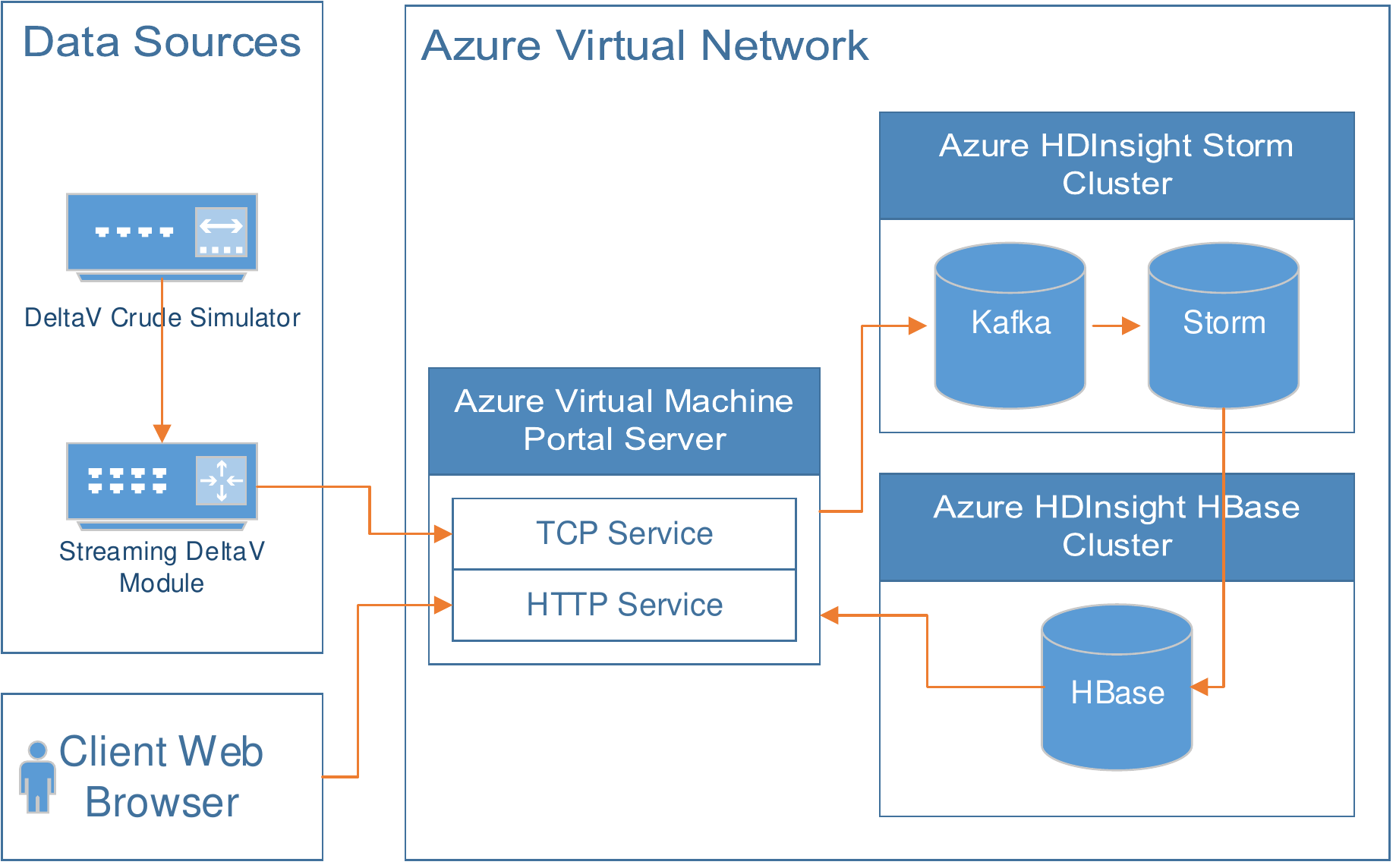}
   \caption{\small System implementation on Microsoft Azure}
   \label{fig:azure-arch}
\vspace{-0.15in}
\end{figure}

Fig.~\ref{fig:azure-arch} presents the system architecture. It follows a Client/Server architecture design. The server provides high-volume data ingest, exactly-once delivery, scalable time-series data storage and real-time parallel data processing. The clients can either push real-time streams into the server or retrieve data (e.g., in the form of queries, visualization, etc.) from the server, through either web or Thrift interface. We created a portal VM to bridge {\rtdap} and external data sources. The portal VM includes 1) a standalone TCP server running on Netty to accept the meta/raw data streams from plant resources using the unified JSON format, 2) a web server to query HBase and provide user-friendly web UI for visualization, and 3) a Storm development tool to define, build and submit Storm topologies for the analytics models developed in the MDS.

A combination of Apache Kafka and Storm frameworks is running in a HDInsight cluster for queuing and real-time processing on data streams received from the portal VM. Raw data were also sent directly to a HDInsight HBase cluster and stored in Azure storage. HDInsight cluster deployment tool provided by Azure allows us to quickly deploy and scale HBase and Storm clusters, while the Kafka cluster needs to be manually installed because it is not yet supported by Azure. Currently we deploy a single node Kafka service in the headnode of Storm HDInsight cluster, and reuse its zookeeper nodes. By default, the HDInsight cluster only exposes the Web manager interface to the Internet which only provides basic and high-level management. We created and deployed the HDInsight Storm cluster, HBase cluster, and the Portal VM in a same Azure virtual network. By doing so, the services provided in the Portal VM can get exposed to the Internet by creating endpoints in the Azure platform and adding firewall exceptions. A connection to a virtual public IP address will be redirected to the Portal VM.

Sample models were created in MDS for fault detection based on both historical data in HBase and real-time data streams flowing through the Storm topology. Power BI and Dashboards were used for reporting, and real-time alerts \& notification. This prototype provides a conceptual validation on the system design. Due to its clean design and potential impact to many industrial sectors, the platform was picked by the Azure team as a case study in UK Azure user group meeting and was discussed in Azure cloud cover show episode 174 on Channel 9~\cite{channel9}.

On the client side, various data sources, either physical plant resources (DeltaV DCS system, wireless Gateways, etc.) or virtual resources (Crude simulator, packet generator, etc.) are connected to {\rtdap} and stream real-time data measurements using a uniformed JSON streaming API. The web interface provided by the web server allows authenticated clients to send queries to {\rtdap} and retrieve both analytics results and raw data from anywhere on any device.

\section{Performance Evaluation} \label{sec:perf}

\eat{
\begin{figure}
   \centering
   \includegraphics[width=0.9\linewidth]{figs/crude.pdf}
   \vspace{0.05in}
   \caption{\small The crude oil refining plant simulator to serve as the data sources in the performance evaluation}
   \label{fig:crude}
\vspace{-0.2in}
\end{figure}}

\begin{figure*}
\hspace{-0.25in}
\centering
	\begin{subfigure}{0.67\columnwidth}
	\includegraphics[width=1.05\linewidth]{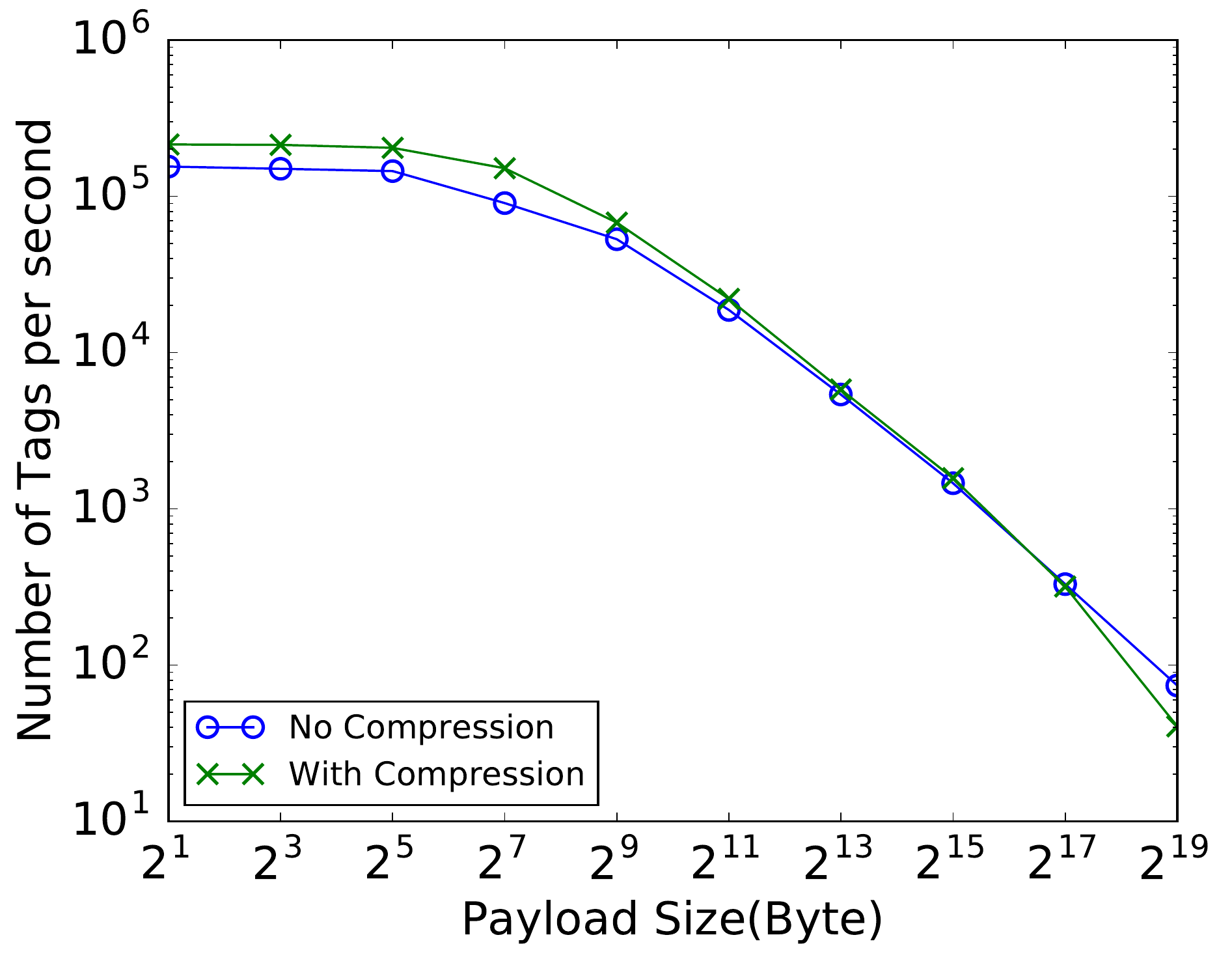}
	\caption{}
	\label{fig:testPortalServerCompression}
	\end{subfigure}
\hfill
	\begin{subfigure}{.67\columnwidth}
	\includegraphics[width=1.05\linewidth]{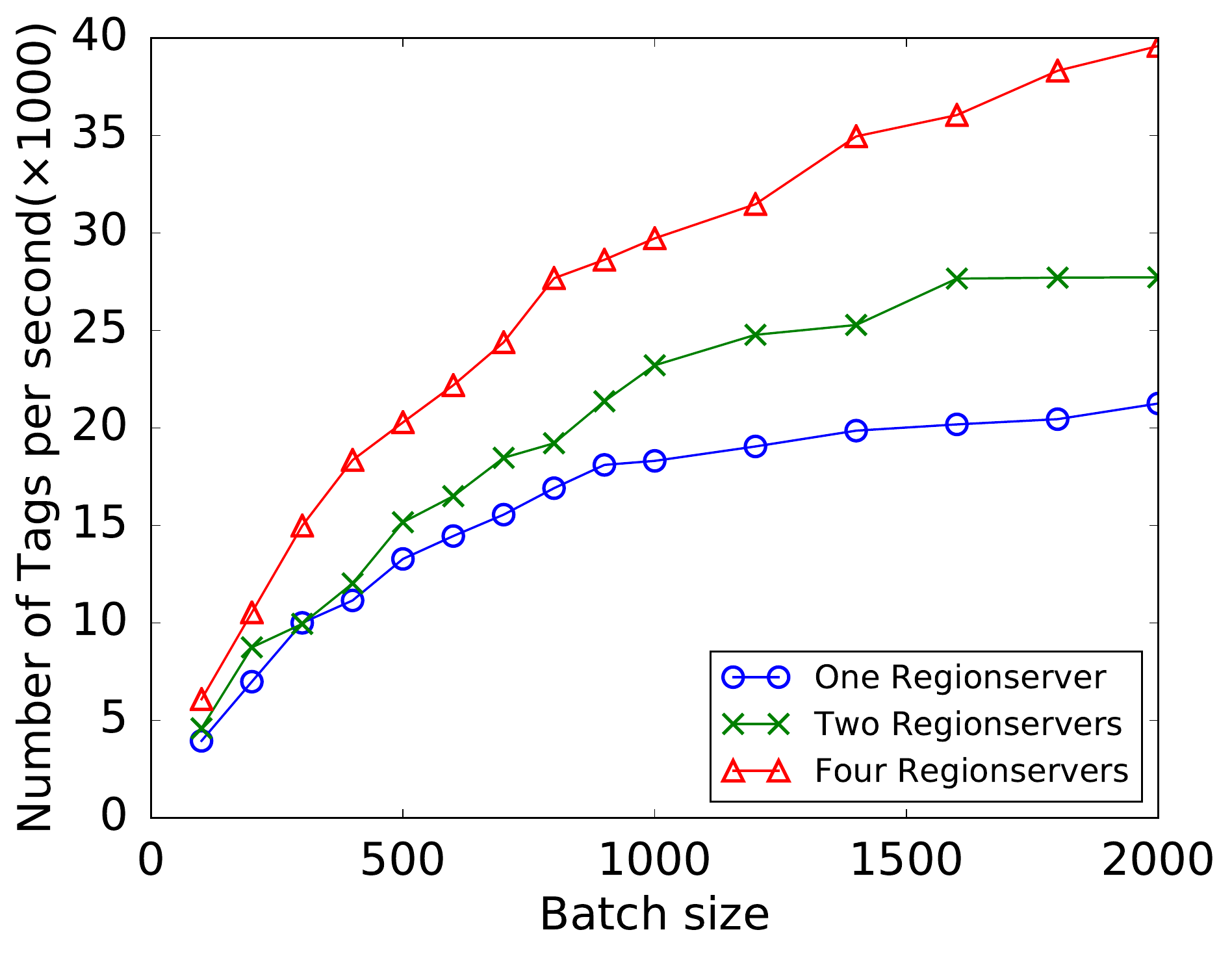}
	\caption{}
	\label{fig:testHBaseWrite}
	\end{subfigure}
\hfill
	\begin{subfigure}{.67\columnwidth}
	\includegraphics[width=1.05\linewidth]{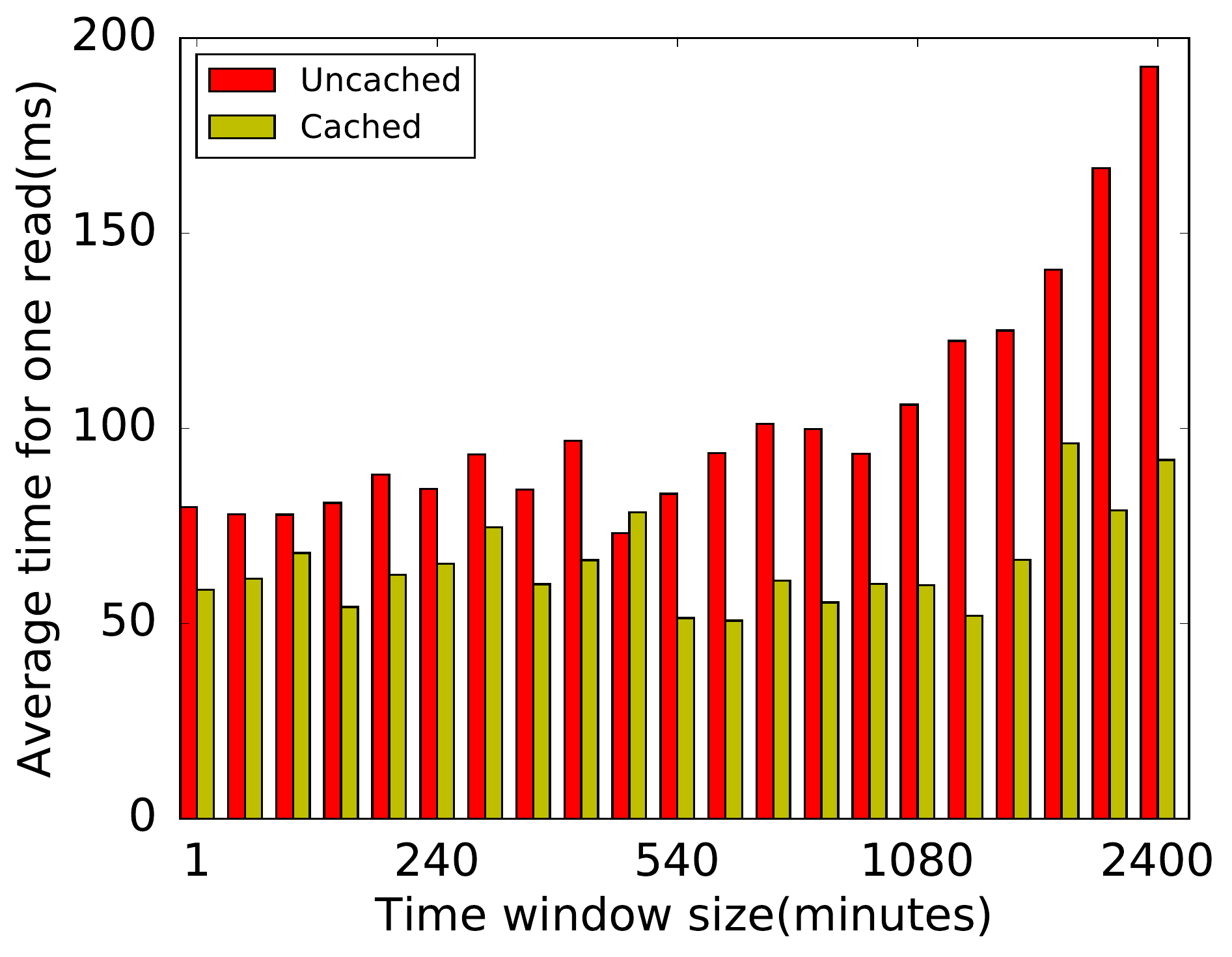}
	\caption{}
	\label{fig:testHBaseRead}
	\end{subfigure}
\vspace{-0.05in}
\caption{\small (a) TCP server throughput w/ and w/o compression mechanism applied on data records; (b) Performance of the write operations; (c) Performance of the read operations}
\label{fig:bigfig}
\vspace{-0.1in}
\end{figure*}

In this section, we evaluate the performance of {\rtdap}. We report our experimental results on the TCP portal server, the time series database, and the runtime execution engine running the aggregation tasks. A crude oil refining plant simulator (Crude simulator for short) is used to be one of the data sources for streaming real-time process measurements into the platform. The Crude simulator can simulate a complete oil refinery process with high fidelity, and all process measurements are accessible via the OPC server.

The key performance metrics used in our experiments are the throughput and latency of the platform in digesting high-volume real-time data streams. For ease of presentation, we implemented a general aggregation function in our performance evaluation, which is a common building block for many batch processing and continuous analytics tasks.

\subsection{Experiment Setup} \label{sec:perf:setup}
Our experiments are performed on the prototype development on Microsoft Azure as described in Section~\ref{sec:impl}. Multiple Crude simulators have been running on workstations to simulate the process measurement flows sent to the real-time data analytics platform from geographically distributed oil refineries over the Internet. A wide range of process measurements are extracted from the Crude simulator through an OPC server and then sent to the portal server via the JSON-based streaming APIs. The tags to be sampled and their corresponding sampling rates are configurable and decided by individual applications. The portal server further forwards the data to different Kafka partitions which are identified by the tag ID, so that the data records for the same tag will always go to the same partition. By doing partitions, we are able to divide work load and parallelize their processing.

\begin{figure}
   \centering
   \vspace{-0.1in}
   \includegraphics[width=0.95\linewidth]{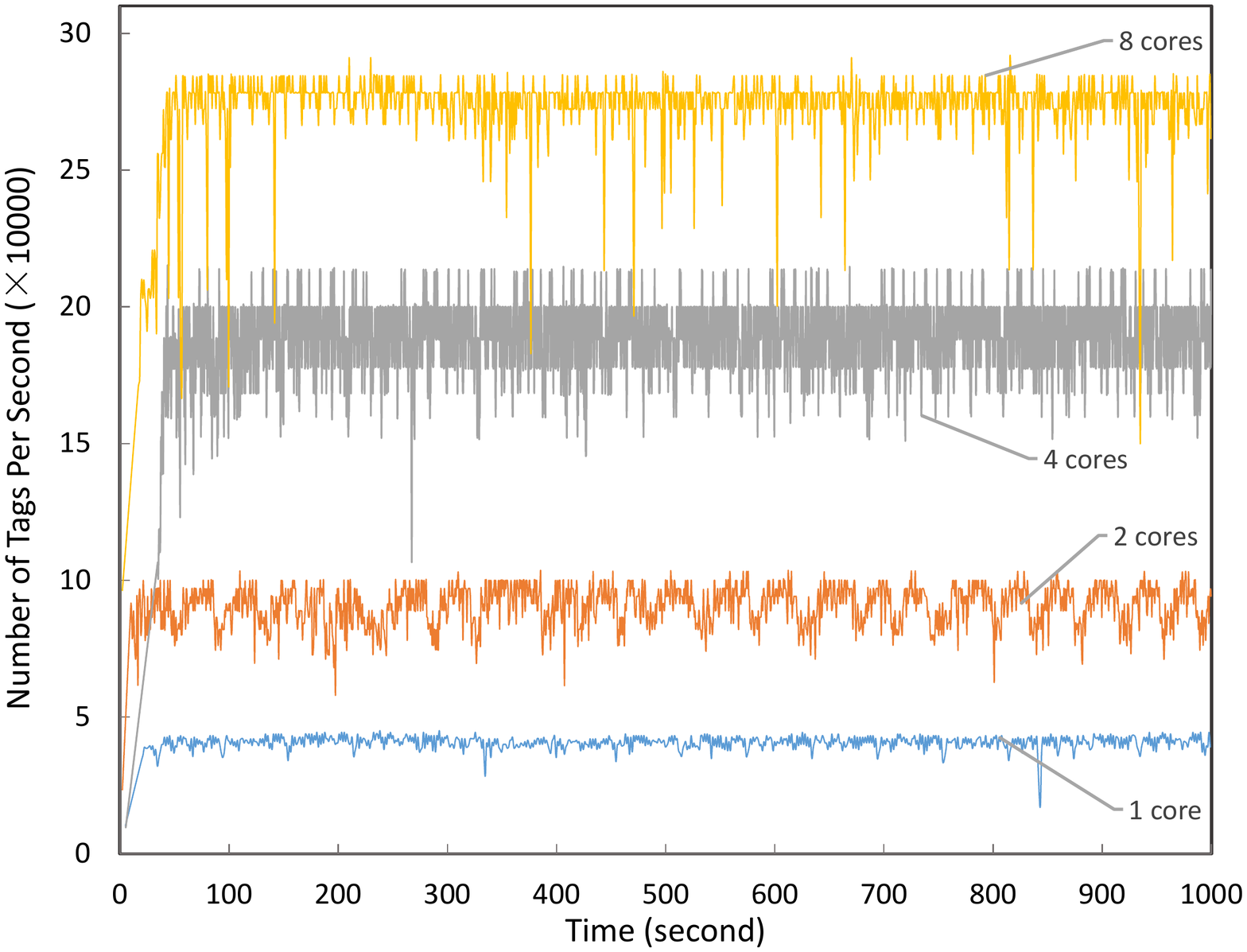}
   \caption{\small Throughput of the TCP server with different computing resources (CPU core number varied from 1 to 8)}
   \label{fig:testPortalServer}
\vspace{-0.2in}
\end{figure}

\subsection{Throughput of the TCP Portal Server}\label{sec:perf:tcp}

In the first set of experiments, we varied the computing resources (in terms of number of cores) on the virtual machine where the TCP portal server is deployed. We evaluated the maximum throughput of the TCP server in terms of number of processed tags per second. It reflects the maximum input throughput of {\rtdap}.

To reach the maximum throughput, instead of using the Crude simulator, we used 7 TCP packet generators installed on different workstations to send JSON objects (in the format of Data Record with a payload size of 70 bytes) to the TCP server at their maximum speeds. The maximum throughput of the TCP server will then be derived when its CPU usage reaches 100\%. As shown in Fig.~\ref{fig:testPortalServer}, we performed four experiments with the number of CPU cores (AMD Opteron 4171 HE) set to be 1, 2, 4, and 8, respectively. Each core is assigned with 1.75 GB memory while all other settings are identical. Each experiment run for 1000 seconds. From Fig.~\ref{fig:testPortalServer}, we observe that when the TCP server run on an one-core machine, it can process approximately $40$K data tags per second. When the number of cores is increased to 2, 4 and 8, the maximum throughput of the TCP sever increases to $90$K, $180$K and $280$K data tags per second, respectively. With this near linear growth of the throughput along with the increased allocation of computing resources, the throughput of the portal server can be easily scaled up.

To evaluate how data compression mechanism affects the throughput of the TCP server, we performed another set of experiments to compare the server throughput by sending compressed and uncompressed data records, respectively. We used the same TCP server as in the last set of experiments and used four cores and 7 GB memory in total. 7 TCP packet generators were used to send JSON objects to the TCP server at their maximum speeds with the payload size varied from 2 bytes to 512K bytes. In the experiments, we used the ``zlib" library~\cite{zlib} for data compression, and the payload was randomly generated using a combination of numbers (0 to 9) and characters. We repeated the experiments to evaluate the throughput with compressed and uncompressed data records.

The experimental results are summarized in Fig.~\ref{fig:testPortalServerCompression}. We have the observation that the throughput of the TCP server is around 200,000 tags per second when the payload size is smaller than 64 bytes. For each tag, once a new data record is received at the TCP server, it took a constant time for processing. A smaller payload size led to a larger number of tags consumed by the TCP server which resulted a higher CPU usage. When the payload size is small, the bottleneck of the TCP server is the CPU resource rather than the network bandwidth. Since it took CPU time to decode the compressed data, applying data compression would downgrade the TCP throughput. On the other hand, when the payload size is larger than 8K bytes, the network bandwidth became the bottleneck. Since a smaller number of tags are transmitted to the server, the CPU resource is sufficient to process all compressed data records. Under this situation, transmitting data records with compression will save network bandwidth which in turn improve the server throughput.

\subsection{Performance of the Time Series Database}

\begin{figure*}
\hspace{-0.25in}
\centering
	\begin{subfigure}{0.67\columnwidth}
	\includegraphics[width=1.05\linewidth]{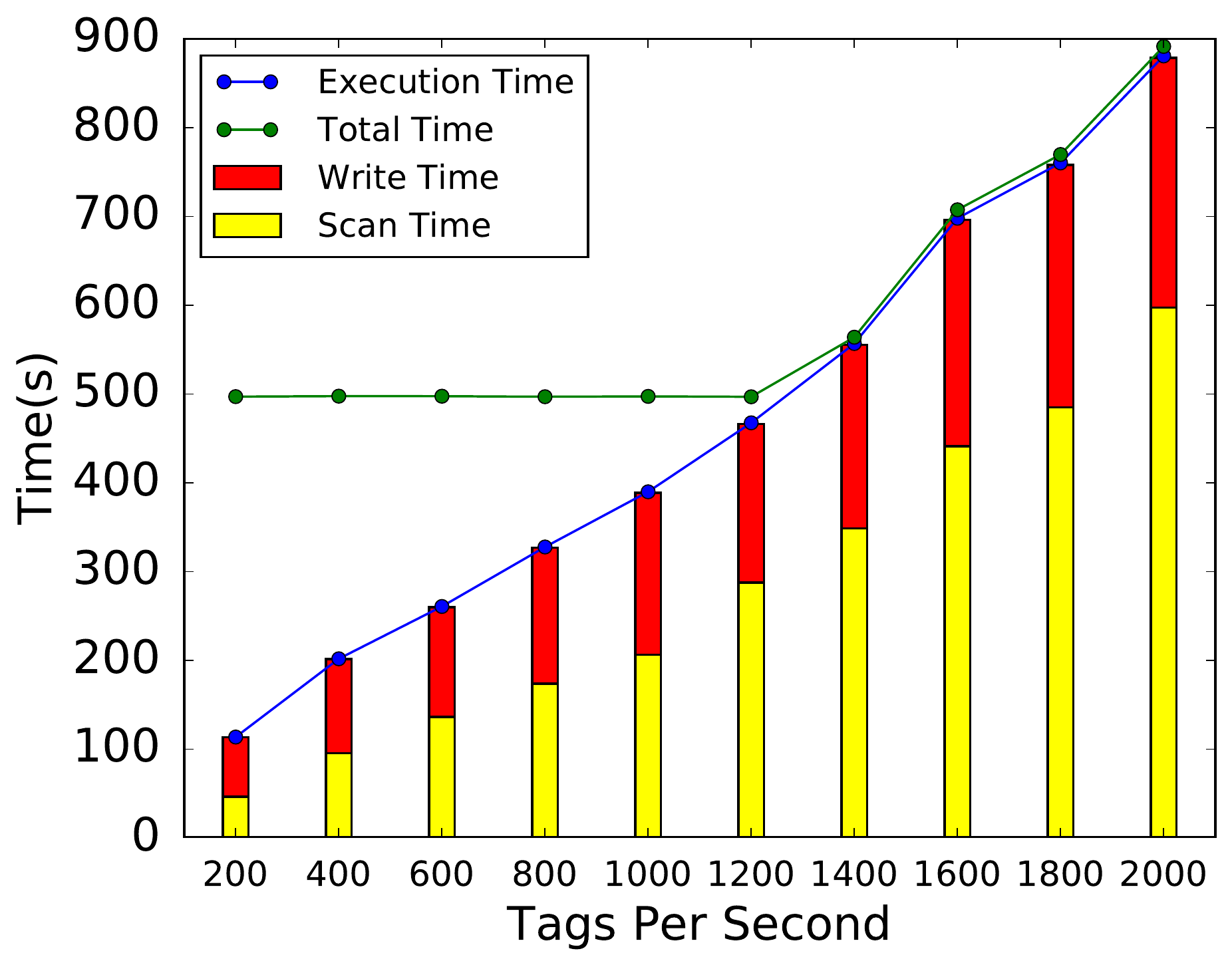}
   \caption{\small }
   \label{fig:ratio}
	\end{subfigure}
\hfill
	\begin{subfigure}{.67\columnwidth}
	\includegraphics[width=1.05\linewidth]{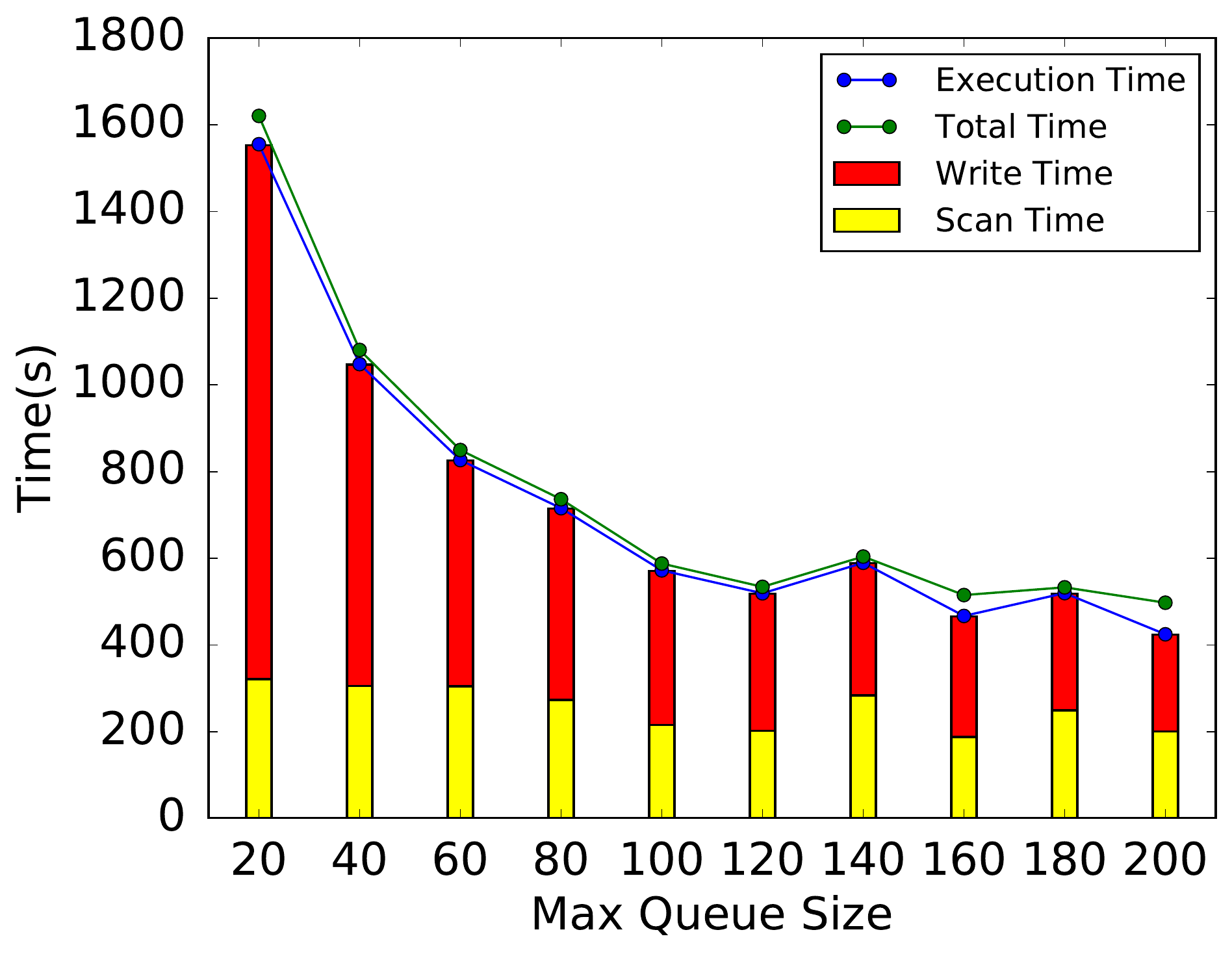}
   \caption{\small }
   \label{fig:max-q}
	\end{subfigure}
\hfill
	\begin{subfigure}{.67\columnwidth}
	\includegraphics[width=1.05\linewidth]{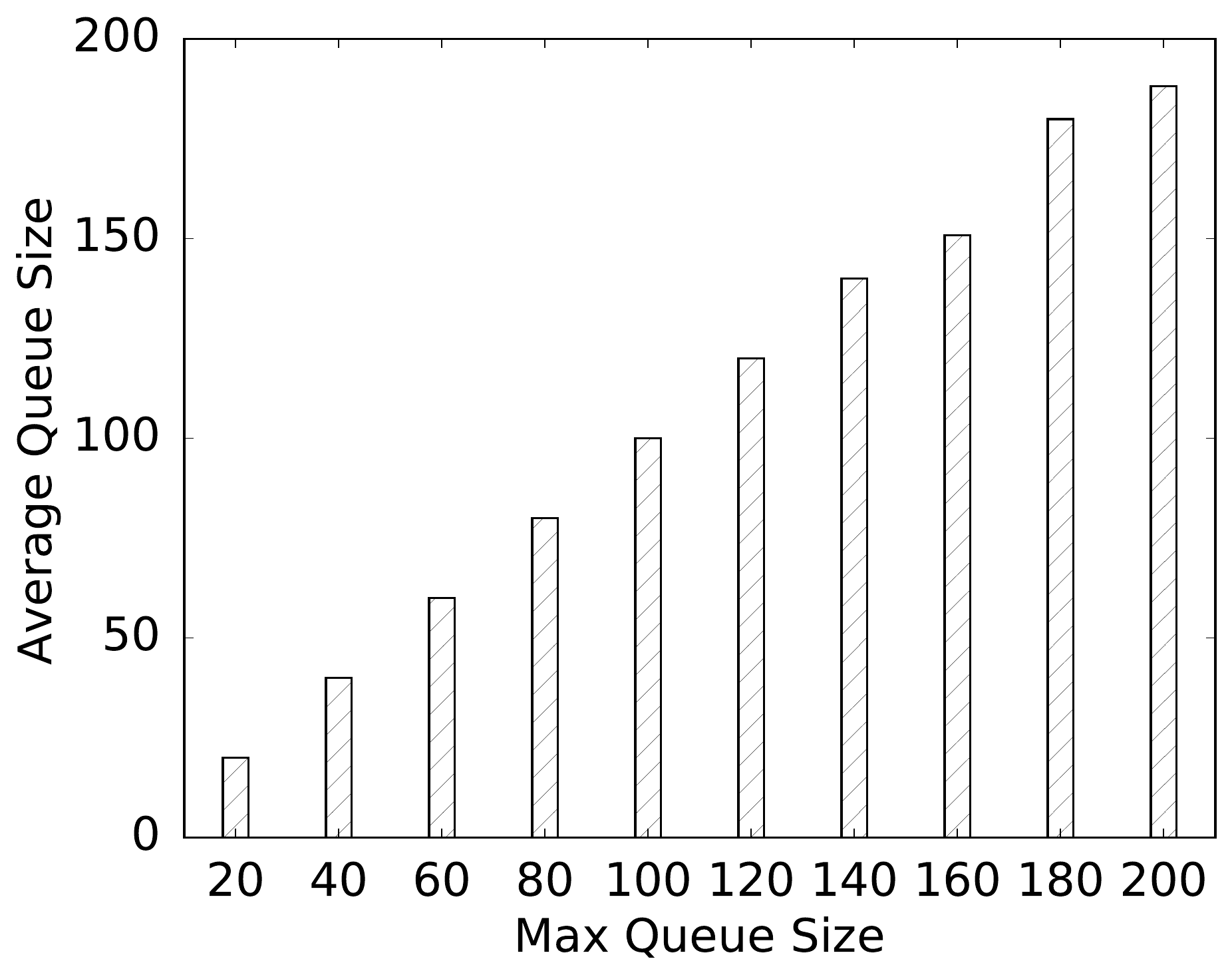}
   \caption{\small }
   \label{fig:ave-q-max-q}
	\end{subfigure}
\vspace{-0.05in}
\caption{\small (a) Distribution of the execution time in the aggregation tasks with varied input speeds; (b) Distribution of the execution time in the aggregation tasks with varied max queue size; (c) Average queue size vs. Maximum queue size}
\label{fig:bigfig2}
\vspace{-0.2in}
\end{figure*}

In order to evaluate the performance of the time-series database schema design, we conducted two sets of experiments to compare the throughput of database write and read under different experimental settings. We tested the write performance by loading an one-month dataset collected from a real-world refinery with a total number of 3,677,625 data records. We tested the read throughput by performing queries on a four-month dataset with a total number of 15,574,062 data records. These data records were loaded from the portal server to the HBase cluster on the Microsoft AZure platform. This cluster comprises of 2 head nodes (one primary and one secondary), one Zookeeper quorum of 3 nodes and a varied number (1-4) of region servers. These machines were configured with the same type of CPU (AMD Opteron 4171 HE) as we chose in Section~\ref{sec:perf:tcp}. HBase head nodes and region servers used 4-core CPUs and the zookeeper masters used 2-core CPUs. We evaluated the read and write performance of the time series database by changing the batch write size and the number of region servers in the HBase cluster. These experiments were also repeated with the HBase caching mechanism enabled and disabled. The experimental results for the write and read performance are summarized in Fig.~\ref{fig:testHBaseWrite} and Fig.~\ref{fig:testHBaseRead}, respectively.

\eat{
\begin{figure}
   \centering
   \includegraphics[width=0.99\linewidth]{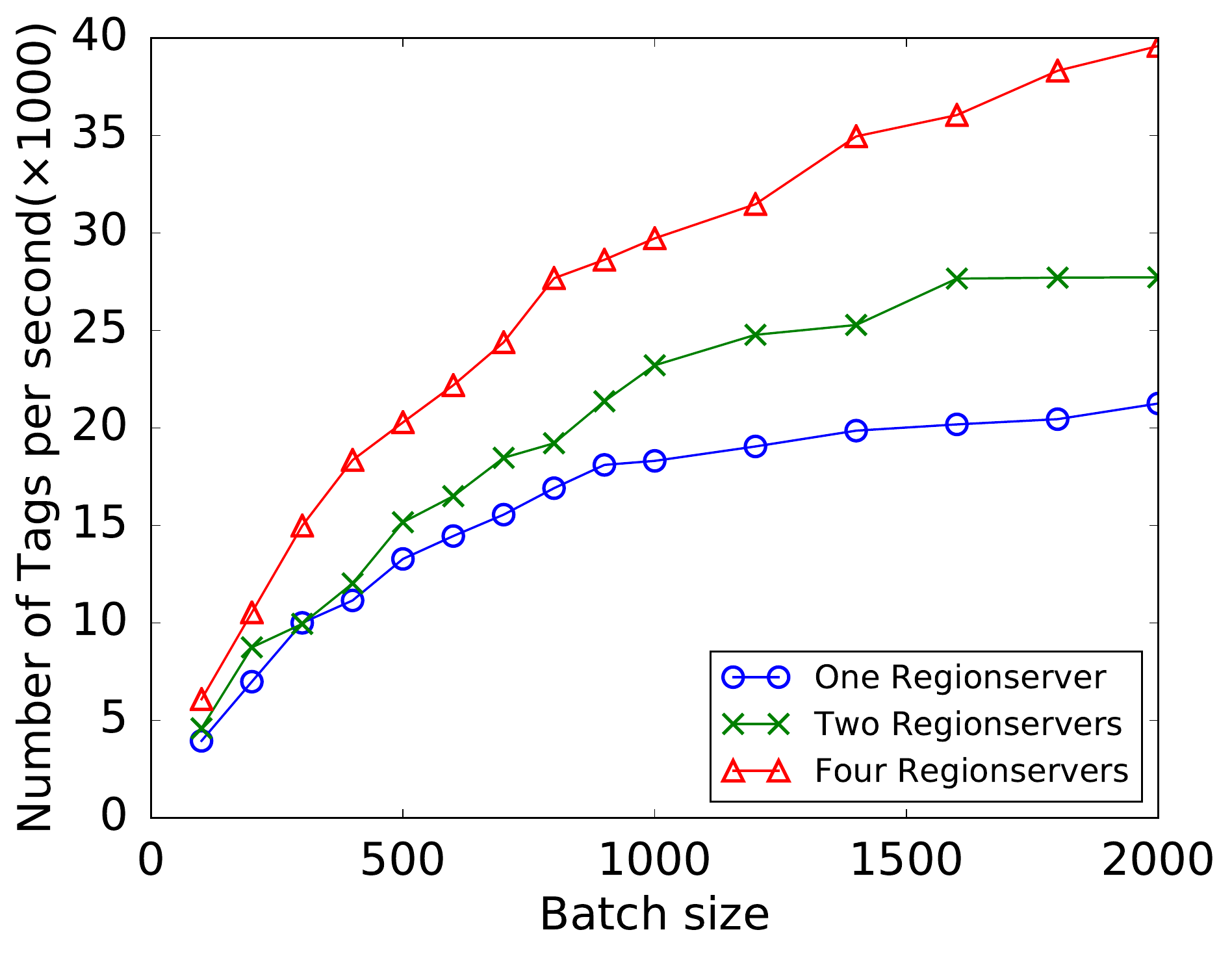}
   \caption{Performance of the write operations}
   \label{fig:testHBaseWrite}
\vspace{-0.2in}
\end{figure}
}

In Fig.~\ref{fig:testHBaseWrite}, we have the observation that the write throughput can be improved with a larger batch size and more region servers. With a batch size of 2000 data records, we can achieve a writing speed of 40,000 data records per second on a cluster with four region servers. In the experiments, we pre-split the raw data table with 1, 2 and 4 regions respectively to ensure that each region server will hold one region. The write requests were distributed almost evenly among the region servers to achieve a better throughput. However, given that the write operations have to hit HBase root and meta tables before they are written in to the raw data table and these operations cannot be paralleled, the write performance does not exhibit a linear improvement.

\eat{
The read tests are performed on a four-month dataset in the raw data table. This table is pre-split with multiple regions according to the number of region servers. In each read test we scan all data records within a time window of one hour and we randomly choose these 1-hour time windows within the four-month data. We perform these read tests 1,000 times and calculate the average execution time per scan. From Fig.~\ref{fig:testHBaseRead}, we observe that each scan operation takes a much longer time than the write operation. This is mainly because HBase will scan the whole region for the specified row key. In addition, in these read tests, we only used one HBase client to submit scan requests. For this reason, we did not observe a performance improvement when the number of region servers increases. Another important technique to accelerate the reading speed is to enable caching.}

The read tests are performed on a four-month dataset in the raw data table.
This table is pre-split with four regions. In each read test we scanned all data records within a time window of varied length from 1 minute to 2400 minutes and we randomly chose these time windows within the four-month time period. We performed these read tests 1000 times and calculated the average execution time per scan. From Fig.~\ref{fig:testHBaseRead}, we observe that the average execution time per scan remains stable when we increased the time window size from 1 minute to 960 minutes. This is consistent with our database schema design, in which each row contains one hour's data records for a given tag. Since the HBase scan is row-based, when the time window is smaller than one hour, we still scan the entire row. From Fig.~\ref{fig:testHBaseRead}, the average scan time is only significantly increased when the size of the time window is larger than 960 minutes. The scan time increased mainly because of the increased sizes of the scan results.

Comparing Fig.~\ref{fig:testHBaseWrite} and Fig.~\ref{fig:testHBaseRead}, we observe that each scan operation took a much longer time than the write operation. This is because HBase will scan the whole region for the specified row key after the scan hits root and meta table. An important technique to accelerate the reading speed is caching. HBase by default enables an LRU cache to accelerate operations on row level.
From the results in Fig.~\ref{fig:testHBaseRead}, we observe that this caching mechanism leads to a doubled scanning speed in average.

\eat{
\begin{figure}
   \centering
   \includegraphics[width=1.0\linewidth]{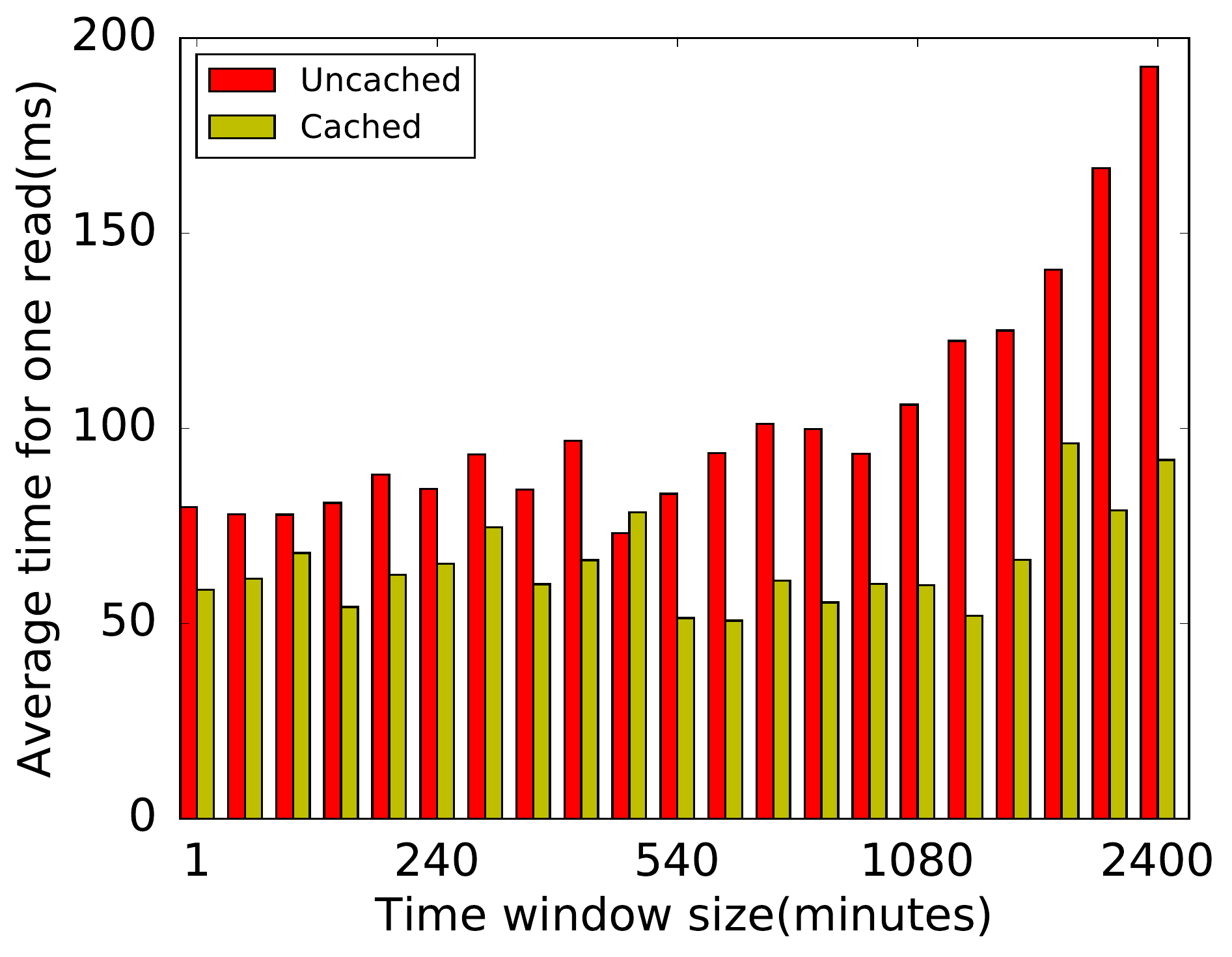}
   \caption{Performance of the read operations}
   \label{fig:testHBaseRead}
\vspace{-0.2in}
\end{figure}
}
\subsection{Performance of the Aggregation Task}

To evaluate the performance of the runtime execution engine, we implemented the time series data aggregation task as described in Section~\ref{sec:design:processing}. Process data flows were streamed into the Kafka via the TCP server. The Spouts in the Storm topology used the Kafka consume function to receive the data records from the associated Kafka partitions, and send them to the corresponding Bolts. The Bolt, once received a data record, first pushed it into the HBase raw data table, and then performed the data aggregation tasks. It first retrieved the aggregated data records from the aggregation tables, calculated the new high, low and close values based on the new received data measurements, and then stored the results back to the HBase at three different time resolutions (minute, hour and day). To complete these aggregation tasks, 3 HBase GET operations and 4 PUT operations are needed.

To reduce the number of HBase accesses during the data aggregation, we implemented a queue in each Spout in order to process multiple data records in batch. To achieve a balance between HBase throughput and latency, we made the queue size self-adaptive. We had one thread on the Spout to keep filling the queue by receiving the data from Kafka. The other thread sent the complete queue to the Bolt and waited until the Bolt finished the aggregation tasks. The size of the queue was affected by the Bolt processing speed. When there were more data records coming in, it took the Bolt a longer time to process, and the Bolt would receive a larger queue in the next round for aggregation. This made the HBase access more efficient but introduced in a larger latency. In the experiments, we bounded the queue size with a maximum number to prevent the scenario when the data input speed is consistently faster than the processing speed of the Storm. Under this situation, the data records will be accumulated in the Kafka, but the Storm will keep a reasonable processing latency, which is the speed of the Bolt processing a maximum size queue of data.

Fig.~\ref{fig:ratio} summarizes the results where we tested the maximum number of tags the aggregation task can process. We let the Crude simulator keep sending data records to the analytics platform for 500 seconds at different speeds, and measured the time it spent on the Bolt to perform the aggregation tasks. In addition to the total processing time, we also measured the HBase scan and write time. In Fig.~\ref{fig:ratio}, the red bar represents the HBase write time while the yellow bar represents the scan time. It shows that the aggregation tasks on the Bolts took negligible amount of computation time when compared to the time spent on HBase access. The results also indicate that when the sending speed was faster than 1200 tags per second, the data could not be processed in time since it would take more than 500 seconds for processing all data records.

\eat{
\begin{figure}
   \centering
   \includegraphics[width=1\linewidth]{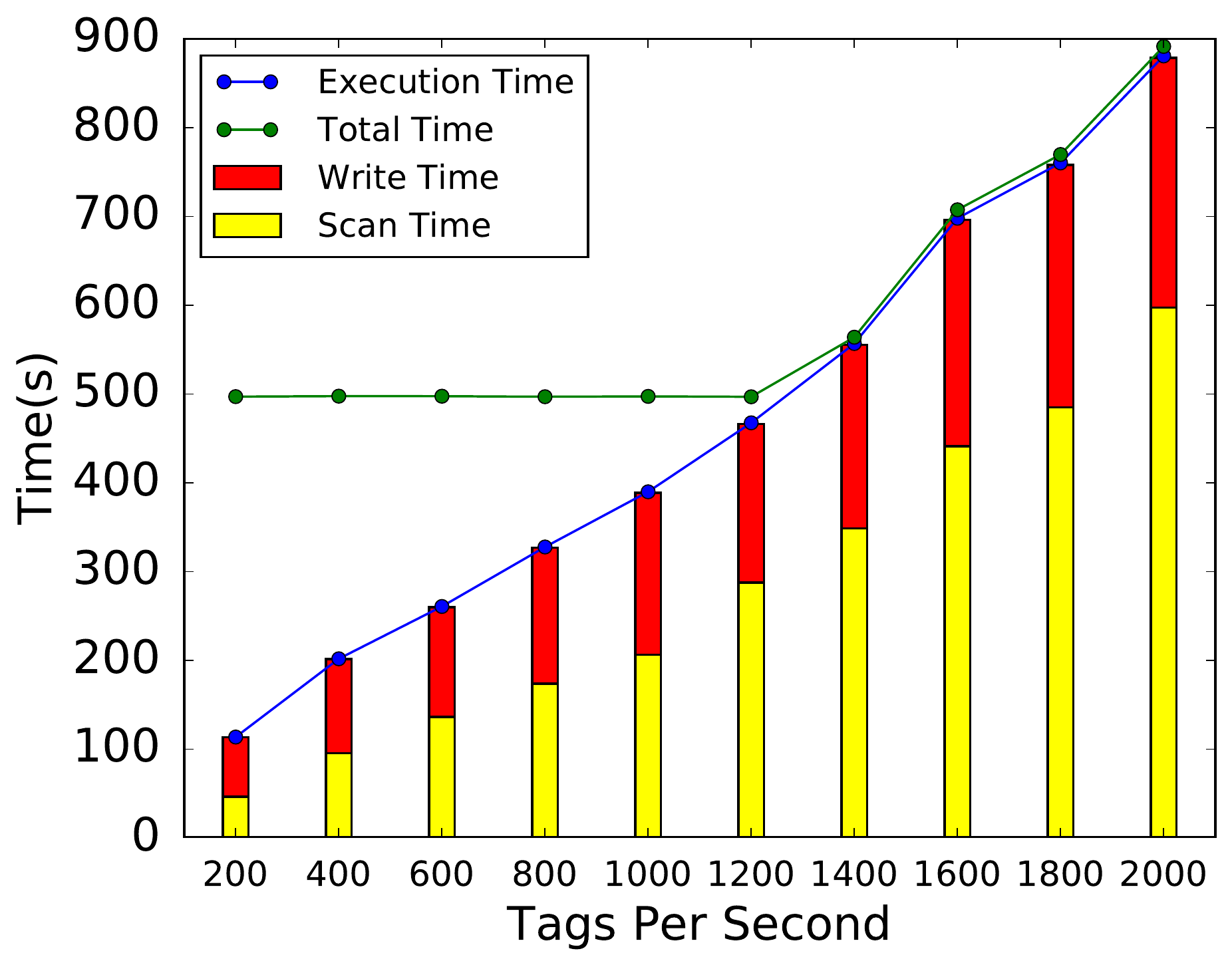}
   \caption{Distribution of the execution time in the aggregation tasks with varied input speeds}
   \label{fig:ratio}
\vspace{-0.2in}
\end{figure}

\begin{figure}
   \centering
   \includegraphics[width=1\linewidth]{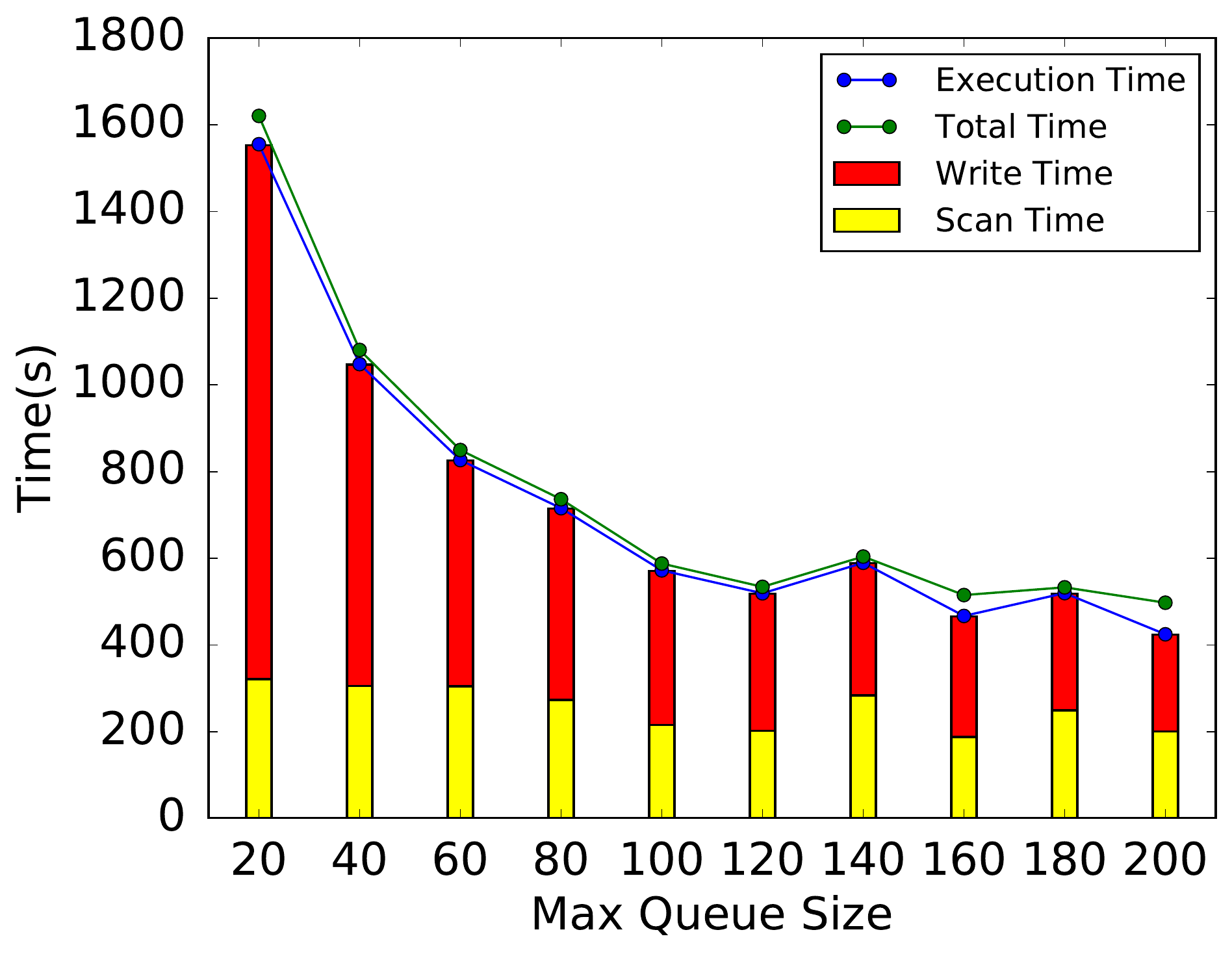}
   \caption{Distribution of the execution time in the aggregation tasks with varied maximum queue size}
   \label{fig:max-q}
\vspace{-0.2in}
\end{figure}
}
Fig.~\ref{fig:max-q} and Fig.~\ref{fig:ave-q-max-q} summarize the results of the experiments where we tested the performance of the aggregation tasks with different maximum queue sizes. In the experiments, we kept the data sources sending data records for a time duration of 500 seconds and fixed the sending speed at 1000 tags per second. We varied the maximum queue size from 20 to 200 and measured the execution times on the Bolts as well as the actual queue sizes in the experiments. In Fig.~\ref{fig:max-q}, we observe that when the maximum queue size increased, the execution time -- especially the HBase write time -- dropped significantly. When the maximum queue size approached to 200, the execution time started to drop below 500 seconds, where the processing speed in the runtime execution engine catched up the sending speed. The similar conclusion can be derived from Fig.~\ref{fig:ave-q-max-q}, where the average queue size in runtime became smaller than the maximum queue size when it went beyond 160.

\eat{
\begin{figure}
   \centering
   \includegraphics[width=0.97\linewidth]{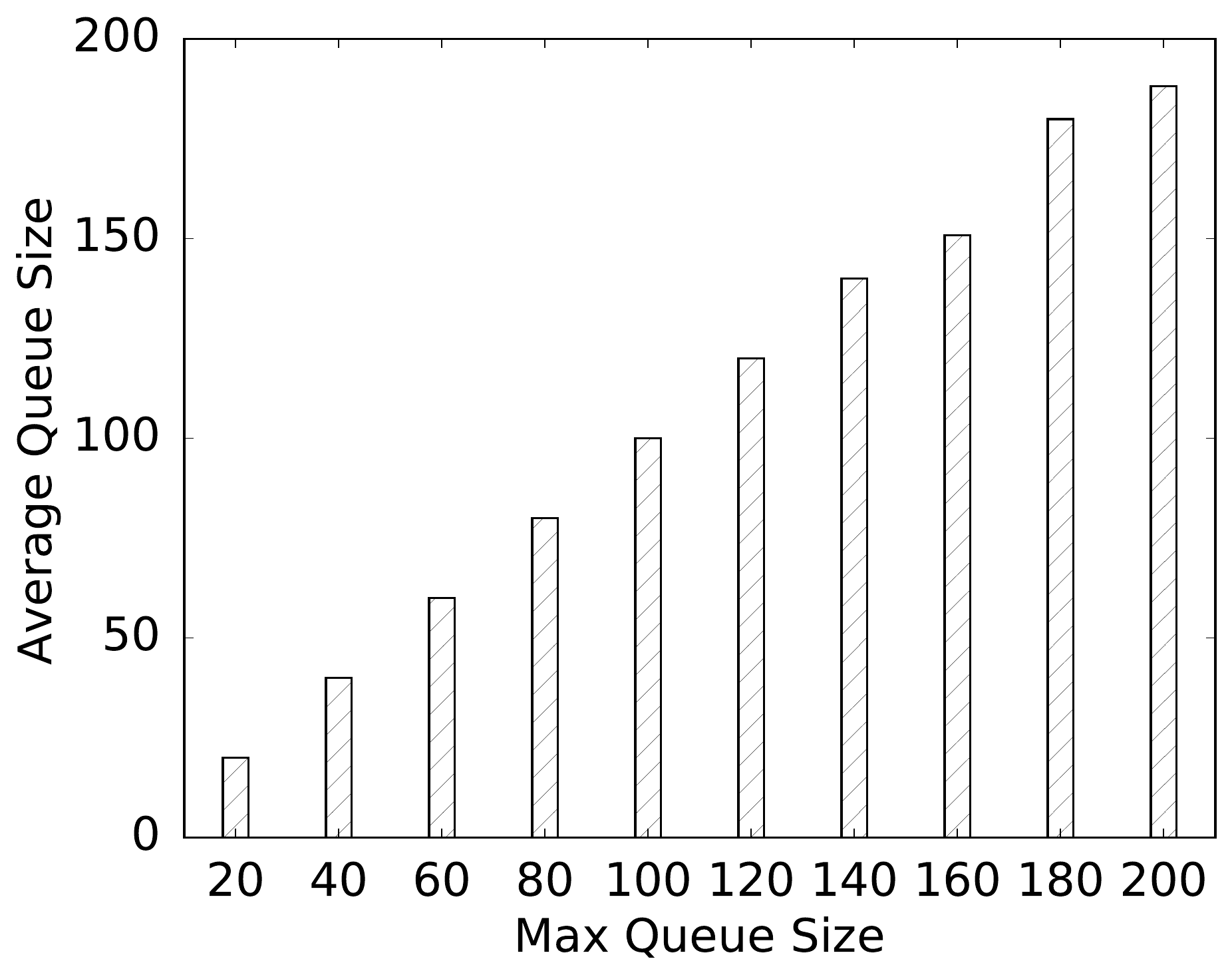}
   \caption{Average queue size vs. Maximum queue size}
   \label{fig:ave-q-max-q}
\vspace{-0.2in}
\end{figure}
}

\section{Conclusion and Future Work} \label{sec:concl}

In this paper, we present the design and implementation of a real-time data analytics platform called {\rtdap} for large-scale industrial process monitoring and control applications. The proposed platform consists of a distributed time-series database for scalable data storage, an analytics model development studio for data and control flow design, and a real-time runtime execution engine to perform parallel and continuous analytics. {\rtdap} can be connected to various plant resources through lightweight industrial IoT field gateway via a unified messaging protocol. Our prototype development on Microsoft Azure and extensive experiments validate the platform design methodologies and demonstrate the efficiency of the data analytics platform in both component and system levels.

For future work, we will extend the time-series database design to support heterogeneous data formats, enhance the real-time parallel processing framework with resource-aware features, and add more analytics models in MDS to support a wider range of continuous analytics tasks. \eat{We envision that the proposed data analytics platform, once mature, will become a paradigm shifter in process industry. In addition, its system architecture and design methodologies can be easily applied to a large array of cyber-physical systems in other industrial sectors, such as healthcare, cybermanufacturing, and building automation.}

\section{Acknowledgement}

The authors would like to thank Joshua Kidd and Noel Bell from Emerson Automation Solutions, and Lara Rubbelke and Wee Hyong Tok from Microsoft   for their great discussions on various technical issues in the design, development and performance evaluation of the {\rtdap} platform. 

\bibliographystyle{IEEEtran}
{\Large \bibliography{all}}

\end{document}